\documentclass[fleqn,usenatbib]{mnras}

\usepackage{newtxtext,newtxmath}

\usepackage[T1]{fontenc}

\DeclareRobustCommand{\VAN}[3]{#2}
\let\VANthebibliography\thebibliography
\def\thebibliography{\DeclareRobustCommand{\VAN}[3]{##3}\VANthebibliography}

\usepackage{scalerel,tikz}
\usetikzlibrary{svg.path}
\definecolor{orcidlogocol}{HTML}{A6CE39}
\tikzset{orcidlogo/.pic={
 \fill[orcidlogocol] svg{M256,128c0,70.7-57.3,128-128,128C57.3,256,0,198.7,0,128C0,57.3,57.3,0,128,0C198.7,0,256,57.3,256,128z};
 \fill[white] svg{M86.3,186.2H70.9V79.1h15.4v48.4V186.2z}
 svg{M108.9,79.1h41.6c39.6,0,57,28.3,57,53.6c0,27.5-21.5,53.6-56.8,53.6h-41.8V79.1z M124.3,172.4h24.5c34.9,0,42.9-26.5,42.9-39.7c0-21.5-13.7-39.7-43.7-39.7h-23.7V172.4z}
 svg{M88.7,56.8c0,5.5-4.5,10.1-10.1,10.1c-5.6,0-10.1-4.6-10.1-10.1c0-5.6,4.5-10.1,10.1-10.1C84.2,46.7,88.7,51.3,88.7,56.8z};
}}
\newcommand\orcidicon[1]{\href{https://orcid.org/#1}{\mbox{\scalerel*{
\begin{tikzpicture}[yscale=-1,transform shape]
\pic{orcidlogo};
\end{tikzpicture}
}{|}}}}

\usepackage{graphicx}	
\usepackage{amsmath}	
\usepackage{siunitx}
\usepackage{mathtools}
\usepackage{float}

\newcommand*{\rom}[1]{\expandafter\romannumeral #1}

\usepackage[normalem]{ulem}

\newcommand{\add}[1]{\textcolor{blue}{#1}}

\newcommand{\hii}[1]{H~\textsc{ii} }
\newcommand{\msun}[1]{$\rm M_{\odot}$}

\newcommand{\aref}[1]{\hyperref[#1]{Appendix~\ref{#1}}}

\title[Stochastic effects on metallicity diagnostics]{Quantifying stochasticity-driven uncertainties in \hii~region metallicities}

\author[]{
Raghav Arora,$^{1,2,4}$\thanks{E-mail: raghav.sofos@gmail.com }
Mark R.~Krumholz$^{\orcidicon{0000-0003-3893-854X}\,1,3}$\thanks{E-mail: mark.krumholz@anu.edu.au}
Christoph Federrath$^{\orcidicon{0000-0002-0706-2306}\,1,3}$\thanks{E-mail: christoph.federrath@anu.edu.au}
\\
$^{1}$Research School of Astronomy and Astrophysics, The Australian National University, Canberra, ACT 2611, Australia\\
$^{2}$BITS, Pilani, Department of Physics, Goa Campus, India\\
$^{3}$ARC Centre of Excellence for Astronomy in Three Dimensions (ASTRO-3D), Canberra, ACT~2611, Australia\\
$^{4}$ Hamburger Sternwarte, Universit\"{a}t Hamburg, Gojenbergsweg 112, 21029 Hamburg, Germany
}

\begin{document}
\setlength{\parskip}{2pt}
\label{firstpage}
\pagerange{\pageref{firstpage}--\pageref{lastpage}}
\maketitle

\begin{abstract}
With the advent of Integral Field Units (IFUs), surveys can now measure metallicities across the discs of nearby galaxies at scales $\lesssim$ 100~pc. At such small scales, many of these regions contain too few stars to fully sample all possible stellar masses and evolutionary states, leading to stochastic fluctuations in the ionising continuum. The impact of these fluctuations on the line diagnostics used to infer galaxy metallicities is poorly understood. In this paper, we quantify this impact for six most commonly-used diagnostics. We generate stochastic stellar populations for galaxy patches with star formation rates varying over a factor of 1000, compute the nebular emission that results when these stars ionise gas at a wide range of densities, metallicities, and determine how much inferred metallicities vary with fluctuations in the driving stellar spectrum. We find that metallicities derived from diagnostics that measure multiple ionisation states of their target elements (e.g.~electron temperature methods) are weakly affected (variation $< 0.1$ dex), but that larger fluctuations ($\sim 0.4$ dex) occur for diagnostics that depend on a single ionisation state. Scatter in the inferred metallicity is generally largest at low star formation rate and metallicity, and is larger for more sensitive observations than for shallower ones. The main cause of the fluctuations is stochastic variation in the ionisation state in the nebula in response to the absence of Wolf-Rayet stars, which dominate the production of $\gtrsim 2-3$ Ryd photons. Our results quantify the trade-off between line brightness and diagnostic accuracy, and can be used to optimise observing strategies for future IFU campaigns.

\end{abstract}




\section{Introduction}
\label{sec:intro}

The gas-phase metallicity of a galaxy is an important parameter for understanding fundamental properties concerning the build-up of chemical elements in galaxies \citep[\add{e.g.,}][]{somerville_primack_1999,nagamine_fukugita_cen_ostriker_2001, lucia_kauffmann_white_2004, dave_finlator_oppenheimer_2011}. However, it is not just the mean metallicity that carries information. The metallicity distribution and gradient within galaxies also carry important information that is sensitive to factors such as the star formation history \citep{chiappini_matteucci_romano_2001}, mass accretion and mass loss via winds \citep{wang_et_al_2019}, and galaxy environment \citep{ kewley_rupke_zahid_geller_barton_2010, rupke_kewley_barnes_2010, rosa_dors_krabbe_hagele_cardaci_pastoriza_rodrigues_winge_2014, torres-flores_scarano_oliveira_mello_amram_plana_2014,philip_chiaki_2017}.
Beyond simple gradients, the azimuthal variation of the metallicity also provides important information regarding various processes that drive gas mixing and inflows, including bar-driven radial mixing  \citep{matteo_haywood_combes_semelin_2013}, spiral-arm driven systematic streaming motions \citep{grand_springel_2016,sanchez_kawata_chemin_perez_2016}, kiloparsec-scale mixing-induced dilutions due to spiral density wave passage \citep{ho_seibert_2017}, thermal and gravitational instabilities \citep{yang_krumholz_2012, petit_krumholz_goldbaum_2015}, and interstellar turbulence \citep{avillez_miguel_mac_2002, krumholz_ting_2018}.

Despite the importance of quantifying the metallicity distribution in galaxies, detailed studies of the azimuthal variations in the metallicity have so far been limited \citep[\add{e.g.,}][]{kreckel_blanc_schinnerer_groves_2016, kreckel_ho_groves_santoro_2019, kreckel20a, vogt_perez_dopita_2017, ho_ting_sharon_2018}. This is mainly due to the observational challenges in measuring nebular emission lines across a substantial portion of the galactic disc with sufficient spatial resolution ($\lesssim$ a few hundred pc). However, with the advent of the wide-field optical integral field unit (IFU) spectrographs, we have gained the ability to obtain high-resolution line emission maps of large samples of nearby galaxies at high resolution. Current IFU-based mapping campaigns include The Physics at High Angular Resolution in Nearby GalaxieS\footnote{http://www.phangs.org
} (PHANGS; \citealt{phangs_survey_2019}), the Calar Alto Legacy Integral Field Area Survey (CALIFA; \citealt{califa_2012}), Mapping Nearby Galaxies at Apache Point Observatory (MaNGA; \citealt{manga_2015}) and the Sydney-AAO Multi-Object Integral-field spectrograph survey (SAMI; \citealt{sami_2015, scott_2018}). The highest resolution of these studies, the PHANGS-MUSE (Multi Unit Spectroscopic Explorer) survey, resolves $<100$~pc scales in tens of nearby galaxies \citep{kreckel_ho_groves_santoro_2019, kreckel20a}. These studies have turned up some interesting anomalies. For example, PHANGS-MUSE reports systematic disagreement between the metallicities inferred using different diagnostics coupled with a scatter of up to 0.3~dex, depending on the diagnostics being compared. This level of uncertainty has the potential to compromise the science we can extract from resolved metallicity maps, and thus it is crucial that we understand its origin. In order to do so, we must understand the assumptions that are baked into metallicity diagnostics.

There are many metallicity diagnostics, each with its own set of limitations and caveats, and with large systematic disagreements ($\sim$~1~dex) amongst them \citep{stasinska_2005, kewley_ellison_2008, peimbert_peimbert_delgado_2017, kewley_nicholls_sutherland_2019}. The most direct methods of measuring metallicity rely on either recombination lines or auroral lines (also known as direct or electron-temperature methods), and these techniques rely primarily on atomic physics, with relatively few additional assumptions or steps. Unfortunately, because the lines they exploit are weak, these methods are often unavailable. For this reason, studies of metallicity beyond the Milky Way and its nearest neighbours often rely on either theoretical or empirically-calibrated diagnostics. The former are derived using models for photoionisation, stellar evolution, and stellar atmospheres, while the latter rely on empirical correlations between the ratios of strong lines and metallicities established by other means.

For both theoretical and strong line methods, one possible source of uncertainty is variation in the shape of the ionising continuum driving the H~\textsc{ii} region. This dependence is clear in the case of theoretical diagnostics, which require the ionising spectrum as an input, and where variations in the shape of that spectrum (for example due to variations in the age of the stellar population, \citealt{byler_dalcanton_conroy_johnson_2017}) can induce corresponding variations in diagnostic line ratios. Even for empirical calibrations, however, variation in spectral shape is a possible concern. While the level of uncertainty in these diagnostics can be assessed by examining the scatter between true metallicity and diagnostic line ratio, and this in principle should include uncertainties due to variation in the spectral shape of the driving source, in practice this method of estimating the uncertainty is valid only to the extent that the sample used to derive the calibration covers the full range of spectral variation that might be encountered. This is generally not the case, since the parent samples used to derive empirical correlations are necessarily biased toward those bright enough to allow determination of the metallicity using auroral lines. As an example, several of the empirical metallicity diagnostics we discuss in this paper have been calibrated from H~\textsc{ii} region samples drawn mainly from the CALIFA survey \citep{califa_2012}, for which the median luminosity in the H$\alpha$ line is $10^{39.0}$ erg s$^{-1}$, and 5th percentile luminosity (i.e., 95\% of the catalogue is brighter than this value) is $10^{37.4}$ erg s$^{-1}$ \citep{Sanchez12b}. By contrast, PHANGS-MUSE includes H~\textsc{ii} regions with luminosities down to $10^{36}$ erg s$^{-1}$ \citep{kreckel_ho_groves_santoro_2019}, and the next generation of IFUs will go even deeper.

One potential source of variation in ionising spectral shape is stochastic sampling of the IMF. While this effect is relatively small for observations of an entire galaxy, with the advent of IFU surveys, this situation has changed. To see why,
a simple back of the envelope estimate suffices. Consider a patch of a galaxy forming stars at a rate per unit area comparable to that of the Solar neighbourhood ($\dot{\Sigma}_* \approx 2.5\times 10^{-3}$ \msun~ Myr$^{-1}$ pc$^{-2}$; \citealt{Fuchs09a}) that is resolved into pixels (or spaxels) that are $100$~pc $\times$ 100~pc in size, comparable to those available in the highest-resolution published studies. The expected star formation rate in the spaxel is $\dot{M}_* = 25$ \msun~ Myr$^{-1}$, and over the $\approx 4$ Myr lifetime of an O star, $\approx 100$  \msun~ worth of new stars should be formed. By contrast, for a star cluster mass function $dN/dM\propto M^{-2}$ from $10^2-10^5$ \msun~, as commonly observed in spiral galaxies \citep{Krumholz19a}, the median cluster mass is $10^{3.5}$~\msun~. This is also the approximate cluster mass required to fully sample the stellar IMF \citep{silva_fumagalli_krumholz_2012}.\footnote{The definition of the mass required to ``fully sample'' the IMF is somewhat arbitrary, since the more rare a particularly stellar mass or type is, the larger the total cluster mass that will be required to sample it. The mass of $10^{3.5}$~\msun~ that we have quoted is obtained using the condition that the expected number of stars with initial mass $>100$ $\rm M_\odot$ exceeds unity.} Thus, a typical 100~pc $\times$ 100~pc galaxy patch similar to the Solar neighborhood has only a $100/10^{3.5} \sim 1$ in 30 chance of containing a cluster young and massive enough to fully sample all the possible ionising stars that go into producing a fully-sampled stellar spectrum. Clearly for such an observation, stochastic variation of the stellar population is a first-order effect that cannot be ignored. We do not expect to approach full sampling until the combination of spaxel size and star formation rate is a factor of $\sim 30$ larger.

Investigations of this issue thus far have been limited. \citet{silva_fumagalli_krumholz_2014} considered the effects of stochastic variation on inferences of the star formation rate from H$\alpha$ in low-luminosity systems, and showed that errors $\gtrsim 1$~dex are common for systems with true star formation rates $\lesssim 10^{-4}$~\msun~yr$^{-1}$. \citet{Krumholz15c} show that a similar-sized stochastic uncertainty affects photometric estimation of the masses and ages of low-mass star clusters, and \citet{Ashworth17a} extended this analysis to studies of the IMF in star clusters. In the most direct precursor to our work here, \citet{paalvast_brinchmann_2017} use Monte Carlo simulations to study how stochastic variation affects determinations of metallicity via nebular lines in dwarf galaxies with low star formation rates. They find significant stochastic errors for galaxies with star formation rates $\lesssim 10^{-3}$~\msun~yr$^{-1}$, an effect they attribute to undersampling of the IMF. In the present work we extend their analysis to the case of IFU observations, using a substantially larger simulation library and considering a larger range of metallicity diagnostics; we also compare the effects of undersampling of the mass distribution and the age distribution of the stellar population and show that the latter is more important than the former, a distinction not made in \citet{paalvast_brinchmann_2017}. Both of these effects contribute when a stellar population does not sample the IMF completely \citep{silva_fumagalli_krumholz_2012}. 


The remainder of this paper is as follows: \autoref{sec:methods} details the method we use to produce a synthetic library of stochastically-sampled stellar populations and corresponding synthetic line emission data. We analyse the properties of this library, and in particular the uncertainties in metallicity inference resulting from stochastic sampling, in \autoref{sec:results}, and discuss their physical origins in \autoref{sect:Disscusion}. We summarise our conclusions in \autoref{sect:conclusion}.

\section{Methods} \label{sec:methods}

Our overall goal in this paper is to characterise 
how much inferred metallicities,
derived from a sample of six commonly used metallicity diagnostics,
fluctuate in response to stochastic variations in the ionising spectrum of the stellar population that drives the \hii~region.
Our method for reaching this goal is to generate a library of ionising spectra from a stochastically-sampled stellar population, predict the nebular line fluxes these would produce, and compare the metallicities derived from these line fluxes to the fully-sampled case and the true inputs. In this section we describe the procedure we use to generate the required library and the metallicity diagnostics that we apply to it.
\autoref{sub_sec:slug_section} describes how we generate our synthetic spectra. \autoref{sub_sec:cloudy_section} describes our physical model of \hii~regions, and how we simulate nebular emission lines within the context of this model. We describe the metallicity diagnostics and their calibrations in  \autoref{sub_sec:Diagnostics}.

\subsection{\textsc{Slug} Simulations}\label{sub_sec:slug_section}

We are interested in generating stochastically-sampled realisations of the ionising spectra produced by stellar populations that reside within a spatial pixel (spaxel), whose size is of order $\sim 100\,\mathrm{pc} \times 100\,\mathrm{pc}$ in a nearby galaxy. For this purpose, we make use of the Stochastically Lighting Up Galaxies code (\textsc{slug}; \citealt{silva_fumagalli_krumholz_2012, silva_fumagalli_krumholz_2014, krumholz_fumagalli_silva_rendahl_parra_2015}), and for the remainder of this section we describe the underlying physical model we adopt, and that \textsc{slug} implements. 

We assume that the average star formation rate in our region of interest has been constant for longer than the lifetime of a massive star, and that stars within our sample pixel form as one or more discrete ``clusters'', each of which can be described as a simple stellar population (SSP). The masses of these clusters are drawn from the observed cluster mass function (CMF), which we parameterise as $dN/dM \propto 1/M^2$ from $M = 10^2 - 10^5$~\msun~ \citep{Krumholz19a}. \citet{Fumagalli11a} show that this hypothesis leads to predicted distributions of H$\alpha$ and FUV luminosity that are in very good agreement with observations. For practical computational purposes we only include in our calculation clusters with ages from 0 to $t_{\rm max}=20$ Myr, since SSPs older than this produce negligible ionising luminosity. The total number of clusters in the stellar population is controlled by the SFR used in the simulation, which is determined by the star formation per unit area and the pixel size. We adopt a fiducial pixel size of $100\times 100\,\mathrm{pc}^2$, and consider star formation rates per unit area $\dot{\Sigma}_* = (0.1-100)\dot{\Sigma}_{*,0}$, where $\dot{\Sigma}_{*,0} =  2.5\times 10^{-3}$~\msun~Myr$^{-1}$~pc$^{-2}$ is the star formation rate in the Solar neighbourhood \citep{Fuchs09a}. Thus we have a range of SFR from 2.5 to 2500~\msun~ Myr$^{-1}$.

We generate our sample stellar populations using \textsc{slug}'s ``Poisson'' sampling method (for full details see \citealt{krumholz_fumagalli_silva_rendahl_parra_2015}). The procedure operates in three steps. We first decide how many clusters to draw. For our chosen CMF, the mean cluster mass is $\langle M \rangle = 690$~\msun~, so the expected number of clusters with age $<t_{\rm max}$ in our sample pixel is $\langle N\rangle = t_{\rm max}\mbox{SFR} / \langle M \rangle$. We draw our number of clusters $N$ from a Poisson distribution with expectation value $\langle N\rangle$. The second step is to assign each cluster an age; since the star formation rate is by assumption constant, all ages are equally likely, and we therefore draw the age of each cluster from a uniform probability distribution from 0 to $t_{\rm max}$. The third and final step is to draw individual stellar masses within each cluster; our procedure for this step mirrors that we use to draw cluster masses. Specifically, we compute the mean stellar mass $\langle m\rangle$ from the IMF; for this paper we adopt a \citet{chabrier_2005} IMF from $0.08 - 120$ $M_\odot$. We then compute the expected number of stars in the cluster, $\langle n_*\rangle = M/\langle m\rangle$, where $M$ is the cluster mass, and draw the actual number of stars $n_*$ from a Poisson distribution with expectation value $\langle n_*\rangle$. Finally, we draw the mass of each individual star from the IMF. At the end of this procedure, we have a complete description of the mass and age of every star in the stellar population.


Our next step is to calculate the spectrum that these stars produce. We compute the physical properties of every star in our population (including the possibility that some stars will already have completed their lives) using the MIST v1 stellar evolution tracks \citep{mist_2016}, and we use \textsc{slug}'s ``starburst99'' stellar atmospheres method to generate the stellar atmospheres and spectra \citep{sb99_1999}; this method uses WM-Basic model atmospheres for O stars \citep{pauldrach_2001}, CMFGen atmospheres for Wolf-Rayet stars \citep{hillier_1998} and \cite{lejeune_1997}, mainly based on Kurucz models, for the remainder of the stellar population . 

We perform $2 \times 10^{5}$ runs for each value of metallicity in the range $\log Z/Z_{\odot} = -1.5$ to 0 in steps of 0.5~dex; here $Z$ is the metallicity normalised to the Solar metallicity ($12 + \log (\rm O/H) = 8.69$). These runs are sampled uniformly in $\log \left(\rm SFR/SFR_{0}\right)$ from $\rm SFR/SFR_{0} = $~0.1 to 100, where $\mbox{SFR}_0 = (2.5\times 10^{-3}\,\textrm{M}_\odot\,\textrm{Myr}^{-1}\,\textrm{pc}^{-2})\times(100\,\textrm{pc})^2$, i.e., $\textrm{SFR}_0$ is the star formation rate of a 100 pc $\times$ 100 pc pixel with a SFR per unit area equal to that of the Solar neighbourhood. We supplement these with an additional $\approx 2.5\times 10^4$ runs at each metallicity with star formation rates sampled uniformly in $\log(\mbox{SFR}/\mbox{SFR}_0)$ over the range $\rm SFR/SFR_{0} = 0.1$--$1$. We require these extra samples at low SFR because a significant fraction of the runs yield line luminosities that fall below the observability cuts we impose below; the additional cases ensure that our statistical errors remain small despite this. Note that our method includes stochastic variation in both the age of the stellar population and the masses of individual stars. As one increases the input SFR, both these distributions become fully sampled.

To provide a baseline for comparison, we also carry out two additional \textsc{slug} runs where we disable part or all of the stochastic sampling. In the first of these, to which we refer below as the fully-sampled case. Note that this is an analytical calculation and is different from the stochastic runs with $\rm M \geq 10^{3.5}$~\msun~ that are approaching a fully-sampled population. We generate a spectrum for a system forming stars continuously at a rate of $0.001$~\msun~yr$^{-1}$ with no stochasticity in either stellar age or mass. That is, in this case we assume that the light is produced by a stellar population containing an infinite number of stars with ages uniformly-distributed from $0$--$20$~Myr, and masses distributed following the \citet{chabrier_2005} IMF; in this case the total star formation rate enters the problem only as a normalisation factor on the overall luminosity. Run in this mode, \textsc{slug} operates identically to a conventional, non-stochastic stellar population synthesis code, and thus the resulting spectrum is derived using the same assumptions usually made in generating theoretical photoionisation models.

In our second comparison case we consider a stellar population that is continuously (i.e., non-stochastically) sampled in stellar mass, but where all the stars have the same age , i.e., where the stars form a single simple stellar population (SSP)
; we view this as being representative of a case where the stellar population is fully-sampled in mass, but not in age. In this case we adopt a stellar population mass of $10^4$~\msun~ and generate spectra for this population at ages from $0$--$10$~Myr in steps of $10^{4}$~yr. Run in this mode, \textsc{slug} acts identically to a conventional stellar population synthesis code simulating the case of star formation in an instantaneous burst. We refer to this case below as the SSP case.

We generate nebular spectra and inferred metallicities for both the fully-sampled and SSP cases using the same procedure as that used for our main library of \textsc{slug} simulations. We describe this procedure next.

\subsection{\textsc{Cloudy} Simulation}\label{sub_sec:cloudy_section}

The radiation from the stellar population is incident on a surrounding H~\textsc{ii} region. We use the C17 version of the \textsc{cloudy} radiative transfer code, last described by \cite{Ferland}, to calculate the resultant nebular spectrum. In our calculation we adopt the photospheric Solar abundances ($Z = 0.0134$, X = $0.7381$) for the nebula for the case $Z = Z_{\odot}$ from \citet{asplund_2009}. For other metallicity values we adjust the abundances of each element using the non-linear and element-specific empirical scaling described in \cite{nicholls_sutherlan_dopita_2017}. 
In addition to abundances, the nebular emission will depend on the geometry and structure of the H~\textsc{ii} region. To specify these choices, we run \textsc{cloudy} in its spherical, isobaric mode, for which the input parameters to the calculation are the inner radius $r_{0}$ and number density $n_{0}$ of the hydrogen nuclei at that radius; \textsc{cloudy} then computes the structure of the remainder of the photoionised region by requiring that the pressure be constant. However, while the parameter $r_0$ is required computationally, it is not necessarily very descriptive of the physical properties of the H~\textsc{ii} region. We therefore choose to define an alternative parameterisation, based on a greatly-simplified H~\textsc{ii} region model, from which we can derive $r_0$.

In our simplified model, the H~\textsc{ii} region surrounding the stellar population is assumed to be isobaric and isothermal, which makes the density uniform as well, i.e., $n_0 = n_{\textsc{ii}}$, where $n_{\textsc{ii}}$ is the mean density. All hydrogen present in the region is assumed to be fully ionised, helium is singly ionised and the radiation pressure is negligible. The H~\textsc{ii} region is a spherical shell, with an inner radius $r_{0}$ and an outer radius $r_{1}$. The inner radius is set by the presence of a bubble of shocked stellar wind material at a temperature $\gtrsim 10^{6}$~K that produces negligible optical or UV emission, and the outer radius by the location where all the ionising photons have been absorbed.

Under the assumptions made above, and further assuming photoionisation equilibrium, the ionising photon luminosity at a distance $r$ from the stellar population located at the centre of the H~\textsc{ii} region is given by \citep{draine_bruce_2011} 
\begin{equation}\label{q_r}
    Q(r) = Q({\rm H}^{0})\left[ 1 - \left( \frac{r}{r_s} \right)^{2} + \left(\frac{r_0}{r_{s}} \right)^{3} \right],
\end{equation}
where $Q({\rm H}^0)$ is the hydrogen-ionising luminosity of the source, defined as  
\begin{equation} \label{q_h}
    Q({\rm H}^{0}) = \int _{0} ^{\lambda_{\rm H^0}} \frac{f_{\lambda}}{h c/\lambda}  d\lambda,
\end{equation}
$\lambda_{\rm H^0} \approx 912$~\AA~is the cut-off wavelength corresponding to the 13.6~eV ionisation potential of H$^0$, $f_{\lambda}$ is the ionising spectrum of the stellar population that is incident on the H~\textsc{ii} region, and $r_{s}$ is the Str{\"o}mgren radius, given by
\begin{equation} 
    r_{s} = \left( \frac{3 Q({\rm H}^{0})}{4\pi \alpha_{B} f_{e} n_{\textsc{ii}} ^{2}} \right)^{1/3}.
\end{equation}
Here $ \alpha_{B} = 2.59 \times 10^{-13}$ \si{cm^{3}\, s^{-1}} is the case B recombination coefficient, $f_{e} = 1.1$ is the abundance of free electrons per H nucleus, and $n_{\textsc{ii}}$ is the number density of hydrogen present in the region. The values of $\alpha_B$ and $f_e$ we have chosen are appropriate for temperatures of $10^{4}$~K and for a region where He is singly ionised. 
Since $r_s$ depends only on $n_{\textsc{ii}}$ and $Q({\rm H}^0)$, \autoref{q_r} allows us to solve for the inner radius $r_0$ in terms of the outer radius $r_1$, defined implicitly by the condition $Q(r_1) = 0$. However, rather than using the outer radius as a our alternative parameter, it is more physically-meaningful to use the wind parameter defined by \cite{yeh_matzner_2012} as  
\begin{equation}
\label{eq:Omega}
    \Omega = \frac{r_{0} ^{3}}{r_{1} ^{3} - r_{0} ^{3}}.
\end{equation}
This quantity is the ratio of the volume occupied by the wind gas to the volume occupied by the ionised gas. Values of $\Omega$ greater than unity signify that winds are dynamically important. For given values of $\Omega$, $n_{\textsc{ii}}$, and $Q({\rm H}^0)$, we can solve for $r_0$ by simultaneously solving the system consisting of \autoref{eq:Omega} and \autoref{q_r} (with $r = r_1$ and $Q(r_1) = 0$). We therefore use $n_{\textsc{ii}} = n_0$ and $\Omega$ as our parameters, in place of $n_0$ and $r_0$.

\citet{yeh_matzner_2012} show that the choice $\Omega \approx 0.3$ produces good fits to the infrared forbidden lines observed in nearly all nearby H~\textsc{ii} regions, so we adopt this value. However, there are a range of densities present in observed H~\textsc{ii} regions, and thus we also treat $n_{\textsc{ii}}$ as a stochastic variable. We draw $n_{\textsc{ii}}$ from a distribution $P(n_{\textsc{ii}})\propto 1/n_{\textsc{ii}}$ (i.e., uniform probability in $\log n_{\textsc{ii}}$) from \mbox{$n_{\textsc{ii}} = 10$--$10^4$~cm$^{-3}$}.

\subsection{Diagnostics}\label{sub_sec:Diagnostics}

The final step in our method is to calculate the gas-phase oxygen abundances ($\log Z \equiv 12 +\rm \log(O/H)$) of H~\textsc{ii} regions. 
For this work, we select six commonly-used optical line diagnostics that span a range from highly accurate but reliant on faint lines, to methods that use only bright lines, but are less accurate. Following the classification scheme used in \cite{kewley_nicholls_sutherland_2019}, we use one auroral line method (the direct $T_{e}$ method), two theoretical optical diagnostics (Kobulnicky and Kewley $R_{23}$ and $\rm D16$), and three empirical diagnostics ($\rm O3N2$, $\rm PP04-N2$ and S-calibration).
Recently there has been some interest in using rest-frame ultraviolet emission lines \citep{Byler20a} for high-redshift galaxies ($z \geq 4$), as the optical lines are expected to be redshifted out of observational windows for ground-based optical and near-infrared facilities. However, we omit these diagnostics because they are still experimental and because the focus of our work is local galaxies. We also omit recombination-line methods, since the lines are very weak and are only detected in nearby Galactic regions. For all of the diagnostics we discuss, our goal is not to provide a comprehensive review, simply to collect the relevant formulae and the uncertainties we use into one place for reader's convenience.
We refer readers to the original papers referenced below for further detail.

\subsubsection{Direct $T_{e}$ Method}

The direct electron temperature ($T_{e}$) method, which was suggested in \cite{gernett_1992}, has been the traditional approach to calculating metallicities. However, this method relies on weak auroral lines, which are rarely detected in galaxies with $12 + \log({\rm O}/{\rm H})> 8.7$ \citep{peimbert_1967, rubin_1969, stasinska_2005}; indeed, in their study of eight nearby galaxies, \cite{kreckel_ho_groves_santoro_2019} report an order of magnitude difference between the number of H~\textsc{ii} regions that have observable auroral lines and the number with detectable optical strong lines. Nonetheless, because this method is the most reliable, and because this method does not rely on making any implicit or explicit assumptions about the shape of the ionising spectrum driving the H~\textsc{ii} region, we start with it to provide a baseline.

The $T_{e}$ method relies on the calculation of the electron temperature $T_e$ and density of the  H~\textsc{ii} region. We first estimate the electron temperature as \citep[their equation 4]{nicholls_2020}
\begin{equation}
    \log t = \frac{7.2939 x + 3.5363}{-0.0074 x^{3} -0.1221 x^{2} + 1.6298 x + 1 },
\end{equation}
where $t = T_e/10^4$~K and 
\begin{equation}
    x = \log_{10}\left (\frac{ [\mathrm{O~\textsc{iii}}] \lambda 4363 }{ [\mathrm{O~\textsc{iii}}] \lambda 5007 + [\mathrm{O~\textsc{iii}}] \lambda 4959 }\right).
\end{equation}
We then use the calculated $T_e$ to estimate the abundances of the first two ionisation stages of oxygen \citep[their equation 3]{izotov_2006}
\begin{equation}
\begin{split}
    12 + \log({\rm O}^{+}/{\rm H}^{+}) = \log \left (\frac{ [{\rm O~\textsc{ii}}] \lambda 3727}{{\rm H} \beta}\right ) + 5.961 + \frac{1.676}{t}\\ {} -0.40 \log t -0.034t + \log(1 + 1.35y),
\end{split}
\end{equation}
\begin{equation}
\begin{split}
    12 + \log({\rm O}^{++}/{\rm H}^{+}) = \log \left (\frac{[{\rm O~\textsc{ii}}] \lambda 4959, 5007}{{\rm H}\beta}\right) + 6.200 \\ +  \frac{1.251}{t} -0.55\log t - 0.014t,
\end{split}
\end{equation}
where $y = 10^{-4} n_{e} t ^{-0.5}$ and $n_{e}$ is the electron density of the H~\textsc{ii} region. Since $n_{e}$ for our models is by construction $\ll 10^{4}$ \si{cm^{3}}, we take $y = 0$. The total oxygen abundance is then calculated as
\begin{equation}\label{te_O_by_H}
    \bigg(\frac{\rm O}{\rm H}\bigg) = \bigg(\frac{\rm O^{+}}{\rm H^{+}}\bigg) + \bigg(\frac{\rm O^{++}}{\rm H^{+}}\bigg),
\end{equation}
neglecting the contribution from the ${\rm O}^{0}$ and the ${\rm O}^{+++}$ ions, which are negligible for H~\textsc{ii} regions produced by massive stars \citep{berg_skillman_henry_2016}. 

The uncertainties in this method arise due to the intrinsic inaccuracies in the atomic data, correction for unseen stages of ionisation, and the collisional cross-section data \citep{kewley_nicholls_sutherland_2019}. While the total level of uncertainty is not known, it is likely small. 

\subsubsection{Kobulnicky and Kewley $R_{23}$}

The Kobulnicky and Kewley $R_{23}$ diagnostic (KKR$_{23}$ hereafter), similar to direct $T_e$, relies on the auroral lines for estimating the metallicity. However, unlike $T_{e}$, this diagnostic is calibrated theoretically on a combination of stellar population synthesis models and photionisation models.  \cite{kewley_dopita_2002} provide an iterative scheme for the R$_{23}$ diagnostic \citep{pagel_edmunds_blackwell_chun_smith_1979}, which accounts for its dependence on the ionisation parameter, and was further improved by \cite{kobulnicky_kewley_2004}. We use the \citeauthor{kobulnicky_kewley_2004} method for our experiments. In this method, we first calculate the ionisation parameter as \citep[their equation 13]{kobulnicky_kewley_2004}
\begin{equation}\label{logq}
    \log q = \frac{32.81 - 1.153y^{2} + a[12 + \log(\rm O/\rm H)]}{4.603 - 0.3119y - 0.163y^{2}+ b[12 + \log(\rm O/\rm H)]},
\end{equation}
where 
\begin{align*}
    a &= 0.1444y^{2} - 0.025y -3.396 , \\
    b &=  0.02037y^{2}  + 0.0271y -0.48, 
\end{align*}
and $y = \log \rm O_{32}$ is defined as
\begin{equation}
    \log \rm O_{32} = \log\left(\frac{[\rm O~\textsc{iii}] \lambda 4959 + [\rm O~\textsc{iii}] \lambda 5007}{[\rm O~\textsc{ii}] \lambda 3727}\right).
\end{equation}

As we can see, \autoref{logq} needs an initial guess for the value of metallicity in the first iteration. For this purpose, we use the input metallicity as the initial guess, because this does not affect the final result, but speeds up convergence. From the ionisation parameter, we estimate the metallicity depending upon the branch into which the initial value falls. For the lower branch, $12 + \log( \rm O/\rm H) < 8.4$, we have \citep[their equations 16 and 17]{kobulnicky_kewley_2004}
\begin{equation}
\begin{split}
    12 + \log(\rm O/\rm H)_\textrm{lower} = 9.40 + 4.65x - 3.17x^{2} - \\ \log q (0.272 + 0.547x - 0.513x^{2}),
    \label{eq:K23_lower}
\end{split}
\end{equation}
while for the upper branch ($\log(\rm O/\rm H) + 12 >8.4$, we have 
\begin{equation}
\begin{split}
    12 + \log(\rm O/ \rm H)_\textrm{upper} = 9.72 - 0.777x - 0.951 x^{2} - 0.072x^{3} \\ - 0.811x^{4} - \log q (0.0737 - 0.0713x - 0.141x^{2} \\ + 0.0373x^{3} - 0.058x^{4}), 
    \label{eq:K23_upper}
\end{split}
\end{equation}
where $x =\log \, \rm R_{23}$ is given by 
\begin{equation}
    \log \,\rm R_{23} = \log \left (\rm\frac{[O~\textsc{ii}]\lambda 3727 + [O ~\textsc{iii}]\lambda 4959, 5007}{\rm H \beta}\right ).
\end{equation}
This yields a new estimate of ${\rm O}/{\rm H}$, which we can in turn use in \autoref{logq} to generate a new estimate of $\log q$, which then yields a new ${\rm O}/{\rm H}$ estimate. We therefore iterate between computing $q$ from \autoref{logq} and ${\rm O}/{\rm H}$ from \autoref{eq:K23_lower} or \autoref{eq:K23_upper} until the values of $q$ and ${\rm O}/{\rm H}$ converge, which usually takes about \mbox{2--3~iterations}.

This diagnostic has been calibrated on some physical assumptions of the H~\textsc{ii} regions - e.g. the dust depletion factor, simplified geometry, uniform composition and isobaric conditions. Since we do not know how much a real  H~\textsc{ii} region will deviate from these assumptions, it is difficult to take this into account in the calibration. 

\subsubsection{S-calibration (Scal)}
\label{subsect: scalOn}
This is a relatively new empirical diagnostic, calibrated by \cite{pilyugin_grebel_2016} from a sample of 313 H~\textsc{ii} regions with $T_e$-based metallicities, with the goal of still correcting for the ionisation parameter as in the $T_e$ and $\mbox{KKR}_{23}$ methods, but using the $\left[\rm S~\textsc{ii}\right] \lambda 6717,6731$ doublet in place of the weak $\left[\rm O~\textsc{ii}\right]\lambda 3727$ auroral line for this purpose -- hence the name S calibration (Scal) -- since the availability of the latter line is often the limiting factor in metallicity measurements using $T_e$ or KKR$_{23}$. Scal has been used by \cite{kreckel_ho_groves_santoro_2019} in place of the $T_{e}$ method because of the unavailability of the $\left [\rm O ~\textsc{ii} \right ] \lambda 3727$ line in MUSE observations.

For this diagnostic, we first use $\rm N~\textsc{ii}$ lines to decide whether a particular $\rm H~\textsc{ii}$ region falls in the upper or the lower branch. For the upper branch, $\log \rm N_{2} > -0.6$, we have \citep[their equation 6]{pilyugin_grebel_2016}
\begin{equation}
\begin{split}
	12 + \log \rm (O/H)_{\rm upper} = 8.424 + 0.030 \log(\rm R_{3}/S_{2}) + 0.751 \log(\rm N_{2}) \\
    + \log(\rm S_{2})\times \left(  -0.349 + 0.182 \log (\rm R_{3}/S_{2}) + 0.508 \log(\rm N_{2})\right),
\end{split}
\end{equation}
and for the lower branch, $\log \rm N_{2} < -0.6$, we have (\cite{pilyugin_grebel_2016} ; their 
\begin{equation}
\begin{split}
	12 + \log \rm (O/H)_{\rm lower} = 8.072 + 0.789 \log(\rm R_{3}/S_{2}) + 0.726 \log(\rm N_{2}) \\
    + \log(\rm S_{2})\times \left(  1.069 - 0.170 \log (\rm R_{3}/S_{2}) + 0.022 \log(\rm N_{2})\right),
\end{split}
\end{equation}
where $\rm N_{2}$, $\rm S_{2}$ and $\rm R_{3}$ are defined as 
\begin{align*}
\rm N_{2} &= \left [\rm N~\textsc{ii}\right ] \lambda 6548, 6584 /\rm H\beta , \\
\rm S_{2} &=  \left [\rm S~\textsc{ii}\right ] \lambda 6717, 6731 /\rm H\beta , \\
\rm R_{3} &=  \left [\rm O~\textsc{iii}\right ] \lambda 4959, 5007 /\rm  H\beta .
\end{align*}

\citeauthor{pilyugin_grebel_2016} find systematic errors of $\leq 0.1$ dex in the Scal metallicities when compared with metallicities derived using the $T_{e}$ method, and their calibration sample covers a relatively large metallicity range, from $\log(\textrm{O/H})\approx 7 - 8.8$.

\subsubsection{$\rm O3N2$}\label{subsect: O3N2}

The O3N2 diagnostic, proposed in \cite{pettini_pagel_2004}, uses the line ratios $\left[\rm O~\textsc{iii}\right] \lambda 5007/\rm H\beta$ and $\left[\rm N ~\textsc{ii}\right]\lambda 6584/\rm H\alpha$. It has been calibrated on the metallicities derived by the direct $ T_{e}$ calculations and photoionisation models. The diagnostic was originally formulated for high-redshift galaxies. However, it has also been used for H~\textsc{ii} regions in nearby galaxies \citep{kreckel_ho_groves_santoro_2019}. It is somewhat less preferred than the other diagnostics since it has been shown to have a high dependence upon the ionisation parameter \citep{kewley_nicholls_sutherland_2019}, but we include it because it is commonly-used in cases where the auroral lines are too faint to be of use. We use the empirical calibration given in \cite{marino_2013}, which the authors derive from a sample of 309 H~\textsc{ii} regions with metallicities obtained using the $T_{e}$ method. The diagnostic is given by \citet[their equation 2]{marino_2013}   
\begin{equation}
    12 + \log(\rm O/\rm H) = 8.533 - 0.214 \times \log\left(\frac{\left[{\rm O}~\textsc{iii}\right]\lambda 5007 /{\rm H}{\beta}}{\left[{\rm N}~\textsc{ii}\right]\lambda 6584/{\rm H}{\alpha}}\right).
\end{equation}
One caveat of this method is the limited sample available to the authors for the calibration, for O3N2 $=$ $-1.1$ to $1.7$, which gives a metallicity range of $ 8.169 \leq 12 +\rm \log(O/H) \leq 8.768$. The authors report a range of $0.18$ dex for $68 \%$ of the observations.

\subsubsection{$\rm PP04-N2$}\label{subsect: PP04-N2}
\cite{pettini_pagel_2004} also propose an optical-line method that just uses the line ratio $\left [\rm N~\textsc{ii}\right ] \lambda 6584/ \rm H{\alpha}$. The availability of the $[\rm N~\textsc{ii}] \lambda 6584$ line in moderate-redshift samples \citep{Byler20a}, especially nearby galaxies, makes it a preferred choice over many other diagnostics. We use the improved calibration of the diagnostic described in \cite{marino_2013}, where they have used $T_{e}$-based metallicities derived from 452 H~\textsc{ii} regions for calibration of the coefficients. The metallicity estimate for this diagnostic is given by  \citep[their equation 4]{marino_2013}
\begin{equation} \label{ppo4_equation}
    12 + \log(\rm O/\rm H) = 0.462 \log\left(\frac{\left[\rm N ~\textsc{ii}\right]}{ \rm H{\alpha}}\right) + 8.743.
\end{equation}

\citeauthor{marino_2013} find that 68\% of the observations lie within 0.16 dex of the $T_{e}$ measurements. This calibration too, has been derived from a limited sample of H~\textsc{ii} regions and thus is only valied for the metallicity range $8.00 \leq 12 + \log(\mathrm{O/H}) \leq 8.65$. For the rest of the metallicity range we use the original calibration of \citet[their equation 1]{pettini_pagel_2004}
\begin{equation}
    12 + \log \rm (O/H) =  0.57 \log\left(\frac{\left[\rm N ~\textsc{ii}\right]}{ \rm H{\alpha}}\right) + 8.90 
\end{equation}

For this metallicity range, 68\% of the measurements have been reported to lie within an interval of 0.18 dex.

\subsubsection{$\rm D16$} \label{subsect: D16}
 
Recently, \cite{dopita_kewley_sutherland_nicholls_2016} proposed a new theoretical diagnostic that uses lines that are not only easy to detect, but are also close to each other so that reddening corrections can also be neglected. The calibration is derived from a grid of photoionisation models run with the \textsc{mappings} code \citep{sutherland_dopita_2018}, which are computed using \textsc{starburst99} \citep{leitherer_schaerer_goldader_delgado_robert_kune_mello_devost_heckman_1999} stellar spectra (identical to what we use here) to describe the shape of the driving radiation field. The lines used in this diagnostic are $\rm H\alpha$, $\left[\rm N~\textsc{ii}\right] \lambda$6584, and the $\left [\rm S ~\textsc{ii}\right]\lambda$6717,31 doublet. Defining 
\begin{equation}
    y = \log\left(\frac{\left [\rm N~\textsc{ii}\right ]}{\left [\rm S~\textsc{ii}\right]}\right) + 0.264 \log\left(\frac{[\textrm{N}~\textsc{ii}]}{\textrm{ H}\alpha}\right), 
\end{equation}
we get the metallicity estimate for this diagnostic, using \citep[their equation 3]{sutherland_dopita_2018}
\begin{equation}
    12 + \log({\rm O}/{\rm H}) = 8.77 + y + 0.45(y + 0.3)^{5}.
\end{equation}

Like other theoretial diagnostics (e.g. KKR$_{23}$), this diagnostic suffers from the limitations of the physical assumptions of the model. \citet{dopita_kewley_sutherland_nicholls_2016} find that the variation in the dust depletion factor is the major source of systematic uncertainties that can be as highs as $\pm 0.12$ dex for non-Solar metallicities. 

\subsubsection{Observability Limits \label{subsect: observability_limits} }

A final consideration that we must apply to all our metallicity diagnostics is observability: especially for low star formation rates, the total ionising luminosity of our stellar population may be low enough that some or all of the lines used in various diagnostics will be unobservably faint. Since we are interested in understanding observations, we must include this effect. Of course the observability limits will depend on details such as the integration time, the size of the aperture, etc., and thus are not the same from one survey to another. Except where otherwise stated, for all the analysis presented here we adopt limits that roughly match the sensitivity of PHANGS-MUSE. Specifically, we only calculate an inferred metallicity for H~\textsc{ii} regions for which the luminosity of the H$\alpha$ line is $>10^{36}$~erg~s$^{-1}$, which is similar to the minimum H$\alpha$ luminosity reported by \cite{kreckel_ho_groves_santoro_2019}.\footnote{It is worth noting that a more accurate treatment of observability limits would be to phrase these limits in terms of line equivalent widths (EWs), rather than absolute luminosities, since in practice what limits the detectability of a line is its brightness relative to the background stellar continuum. However, our modelling pipeline does not enable us to calculate the EWs and so we use the absolute line luminosity as a rough proxy.} We also require that the other lines used in diagnostics be bright enough relative to this so that they too will be observable. To this end, we adopt a minimum line luminosity of $10^{-1.5}$ times the H$\beta$ flux for oxygen lines, comparable to the limiting line ratios found in \citet{pilyugin_grebel_mattson_2012} and \citet{kreckel_ho_groves_santoro_2019}, and $10^{-2.5}$ times the H$\alpha$ line for nitrogen lines, comparable to the limiting line ratio found in \citet{marino_2013}. If one of our sample H~\textsc{ii} regions has a line luminosity below our observability cut, we do not compute any of the metallicity diagnostics that rely on that line.

There are some important implications of our choice of sensitivity limit. One is that, at low SFRs, the sample of H~\textsc{ii} regions for which we derive an inferred metallicity is not the same for all diagnostics. For example, in our lowest SFR bin (see below), roughly five times as many H~\textsc{ii} regions have detectable [N~\textsc{ii}] $\lambda6584$ lines (and thus can have their metallicities inferred using the D16 diagnostic) as have detectable [O~\textsc{iii}] $\lambda4364$ lines (as required for KKR$_{23}$ or $T_e$). Compared to the cases with detectable [N~\textsc{ii}] $\lambda6584$, those that also have detectable [O~\textsc{iii}] $\lambda4364$ are present in cases where the stochastic draw happened to produce particularly massive stars, or stars at stages of stellar evolution where they produce the hard photons needed to produce O~\textsc{iii} emission. Secondly, we also observe the well-known effect \citep{peimbert_1967, rubin_1969, satinska_2005} that the number of H~\textsc{ii} regions that have an observable [O~\textsc{iii}] $\lambda4364$ line falls tremendously at high metallicities. This effect is also seen for the [N~\textsc{ii}] $\lambda6584$ line at low metallicities. We emphasise that these are real physical effects, that can potentially account for some of the observed systematic differences between metallicities inferred with different diagnostics. We therefore make no attempt to correct for or remove these effects in our analysis.

\begin{figure}
    \centering
    \includegraphics[width=0.47\textwidth]{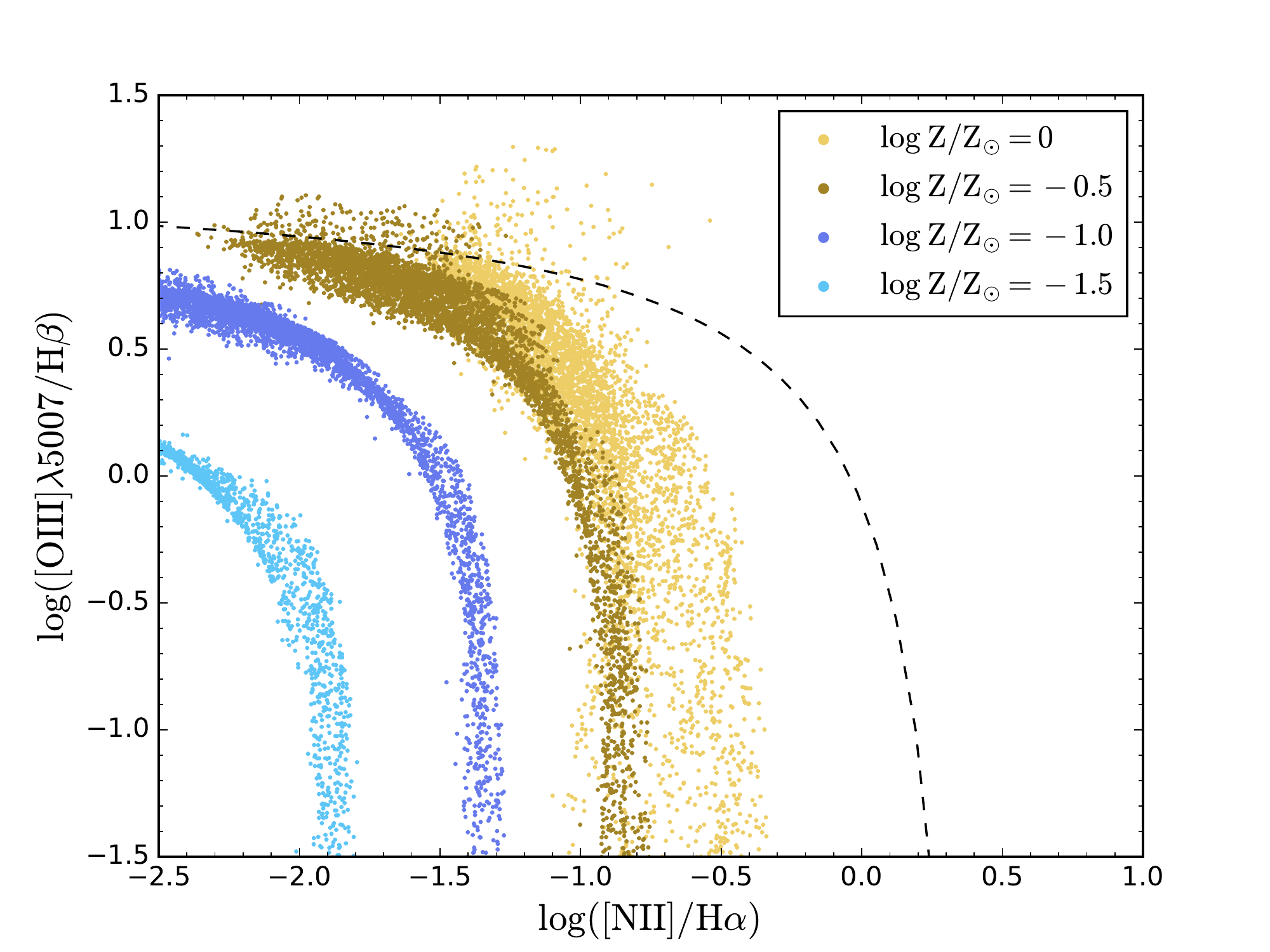}
    \caption{
    The BPT diagrams of our models after applying the observational cuts of $\rm H\beta>10^{36}$~\si{ergs/s}, $\rm [O~\textsc{iii}\rm] \lambda 5007/H{\beta} > 10^{-1.5}$ and $\rm [N~\textsc{ii}\rm] \lambda 6584/H{\alpha}> 10^{-2.5}$; to make the plot readable, we show only $\approx 5\%$ of our realisations. The four different colours show four different values of the input metallicity, as indicated in the legend.
    The dotted line represents the extreme starburst line from \citet{kewley_2001}. Only a small fraction of $\rm H~\textsc{ii}$ regions cross over to the AGN region above the dotted line.}
    \label{fig:BPT}
\end{figure}

In order to verify that our sampled H~\textsc{ii} regions have reasonable line luminosities, and that they would not be mistaken for AGN in an observed sample, in \autoref{fig:BPT} we plot the BPT diagram for our simulation library. This diagram includes only those H~\textsc{ii} regions that pass our observability screen, as described above. We see that the locus occupied by the H~\textsc{ii} regions in this diagram is below the \citet{kewley_2001} starburst line with a few exceptions, and generally populates a region consistent with observed H~\textsc{ii} regions.

\section{Results} \label{sec:results}

Our library of H~\textsc{ii} regions models varies with three input parameters:
\begin{itemize}
\item Star formation rate (SFR), which varies continuously from $0.1$-- $100\times\mbox{SFR}_0$, where $\mbox{SFR}_0 = 25$~\msun~~Myr$^{-1}$ is our fiducial SFR for a $100\times 100\,\mathrm{pc}^2$ spaxel forming stars at the same rate as the Solar neighbourhood.
\item  Metallicity, which takes on four discrete values, $\log Z/Z_\odot = -1.5$, $-1.0$, $-0.5$, and $0$.
\item Ionised gas number density, which varies continuously from $10$--$10^4$~cm$^{-3}$. 
\end{itemize}
For each model case, our output consists of the metallicity that we would infer from each of our six sample line diagnostics at 20~Myr from the time our model began forming stars. In the following sections, we discuss how the level of stochastic fluctuation in inferred metallicity varies in response to changes in the star formation rate, true metallicity, and survey sensitivity.
For the purposes of this analysis we limit our sample to H~\textsc{ii} regions within which the ionised gas number density $n_\textsc{ii} = 10$--$100 \,\si{cm^{-3}}$, the range most commonly observed for optically-selected $\rm H~\textsc{ii}$ regions \citep{tremblin_2014}. We show the effect of having an ionised gas number density far from the commonly observed values in \aref{appendix}.

\subsection{Variation with Star Formation Rate and Metallicity}\label{subsect: Star Formation Rate}

\begin{figure*}
    \hfill
     \centering
     \includegraphics[width =\textwidth]{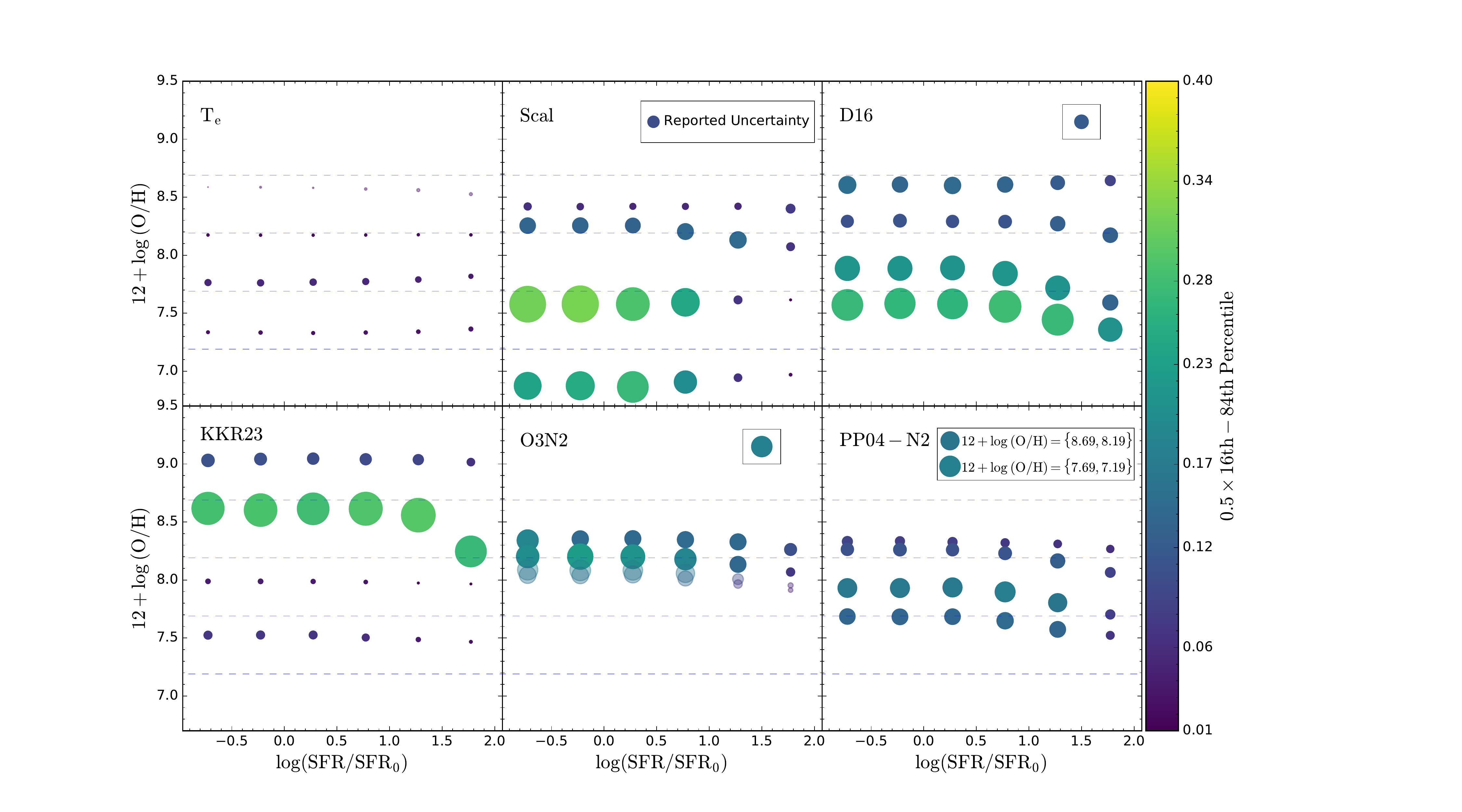}
     \caption{Each of the six panels show the median and half of the 16th to 84th percentile range (equivalent to $1\sigma$ dispersion for a Gaussian distribution) in metallicity $12+\log(\mathrm{O/H})$ derived from the indicated diagnostic. We show these as a function of star formation rate, $\rm SFR/SFR_{0}$ (horizontal axis), and true (input) metallicity (vertical axis). The center of each point in the figure indicates the median of the derived metallicity for all the runs in the given SFR and true metallicity bin. The true metallicity bin is indicated by the relative placement of the point on the vertical axis, starting from the bottom with the lowest metallicity bin to the top representing the highest metallicity bin respectively. The colour and size of the point indicates the size of 16th to 84th percentile spread in that bin. The reported uncertainties, wherever available, are indicated in the legends of each panel. Faded points represent the range in which the diagnostic is not reliable - either due to its limited calibration or because we did not have significant number of \hii~ regions that passed our observational cuts in that particular bin. In all cases the runs shown are those with H~\textsc{ii} region densities $n_{\textsc{ii}} = 10$--$100$~cm$^{-3}$.
     }
     \label{fig:variation_with_SFR_metallicity}
 \end{figure*}

We begin by examining the effects of stochasticity at different metallicities and star formation rates (SFRs). To this end, we select from our library all models with a particular discrete metallicity  and density $n_{\textsc{ii}} = 10$--$100$~cm$^{-3}$, and bin the models in $\mbox{SFR}/\mbox{SFR}_0$; our bins are $0.5$~dex wide, and run from $\mbox{SFR}/\mbox{SFR}_0 = 0.1$ to $100$. For each bin, we characterise the level of stochastic fluctuation in the inferred metallicity by calculating half the 16th to 84th percentile range (equivalent to the $1\sigma$ disperison for a Gaussian distribution) in metallicity, $12 + \log(\mathrm{O/H})$, derived from each diagnostic, including in this calculation only those H~\textsc{ii} regions in our sample library that are bright enough to pass our observability cuts. We show the result in \autoref{fig:variation_with_SFR_metallicity}. Each panel of \autoref{fig:variation_with_SFR_metallicity} is for a particular diagnostic, and within each panel, the position along the vertical axis shown by dotted lines, marks the true input metallicity of our models and position along the horizontal axis shows $\mbox{SFR}/\mbox{SFR}_0$. The center of each point lies on the median of the metallicity values and the size of the point and color indicates the spread in the metallicity values, as indicated in the colorbar, given by a diagnostic at a particular input metallicity value and bin of $\mbox{SFR}/\mbox{SFR}_0$.

First, we see that the metallicity, as derived from different diagnostics, have their own systematic offsets. These have been studied before in detail \citep{stasinska_2005, kewley_ellison_2008, peimbert_peimbert_delgado_2017, kewley_nicholls_sutherland_2019}. Unless stated otherwise, we will be focusing on the uncertainties of these diagnostics. We see that, as expected, there are almost no fluctuations in the metallicity derived from the $T_e$ method, independent of the star formation rate or metallicity. This is as expected: the $T_e$ method is based on fundamental atomic physics, and relies on essentially no assumptions about the shape of the ionising spectrum. Thus this diagnostic offers a useful control for our experiment.

For all other diagnostics we see a generic trend that, at almost all true metallicities, the dispersion in the inferred metallicity increases as we lower the SFR, though the amount of increase varies by diagnostic and by true metallicity -- for example, KKR$_{23}$ shows only a weak dependence on SFR, whereas D16 and Scal show a very strong dependence, particularly at low metallicity; the others are intermediate.

We can interpret the results as follows. A larger value of $\rm SFR/SFR_{0}$ simply entails more stars in our fiducial patch and a better sampling of the IMF and stellar ages. A value of $\rm SFR/SFR_{0}$ $\sim$ 100, on average, will give us $\approx 10^{4}$~M$_{\odot}$ worth of stars in 4~Myr, more than the $\sim 10^{3.5}$~\msun~ that we take to be the mass required for a full sampling of the IMF \citep{silva_fumagalli_krumholz_2012}. Thus, for our highest SFR bin, the 16th to 84th percentile range is in almost all cases $<0.1$ dex. (The primary exception to this statement is KKR$_{23}$ at a true metallicity of $12 + \log(\textrm{O/H}) = 8.19$, which has a large dispersion because this metallicity lies close to the break between the upper and lower branches of the diagnostic.) Stochasticity has little effect in this limit, and the uncertainty budget is instead dominated by other sources.

By contrast, for $ \rm SFR/SFR_{0} <10$, the expected mass is $<10^3$~M$_\odot$, and we are far from the fully-sampled regime. Depending upon the diagnostic and the true metallicity value, this can induce large or small fluctuations in the inferred metallicity. In the worst cases, e.g., Scal at low SFR and metallicity, the dispersion can be $\sim 0.3$ dex, substantially greater than the $\approx 0.1-0.15$ dex nominal uncertainty. We emphasise that these large dispersions, even for empirically-calibrated diagnostics like Scal, are not surprising in light of our discussion in \autoref{sec:intro}: the calibration sample for Scal comes from CALIFA, and therefore consists of H~\textsc{ii} regions $\gtrsim 30$ times more luminous than the sensitivity limit that we are now exploring. Stochasticity is expected to have a larger effect in this regime.

\subsection{Variation with Survey Sensitivity}\label{{subsec: variation_with_sensitivity}}

Our fiducial choice of sensitivity limit is typical of modern IFU surveys such as PHANGS-MUSE (\autoref{subsect: observability_limits}). However, it is also of interest to ask whether we expect to find similar results for a deeper survey, as might be possible with a next-generation IFU on a 20-30m class telescope, e.g., GMTIFS on the GMT or HARMONI on the E-ELT. To this end, we repeat our analysis in the previous section, but using an absolute cut-off on the H$\alpha$ line luminosity of $10^{35}$ erg s$^{-1}$, i.e., a factor of 10 deeper than current limits; we do not change the observability cutoffs for other lines, which are expressed as brightness relative to the H$\beta$ line, rather than absolute luminosity.

\begin{figure*}
    \hfill
     \centering
     \includegraphics[width =\textwidth]{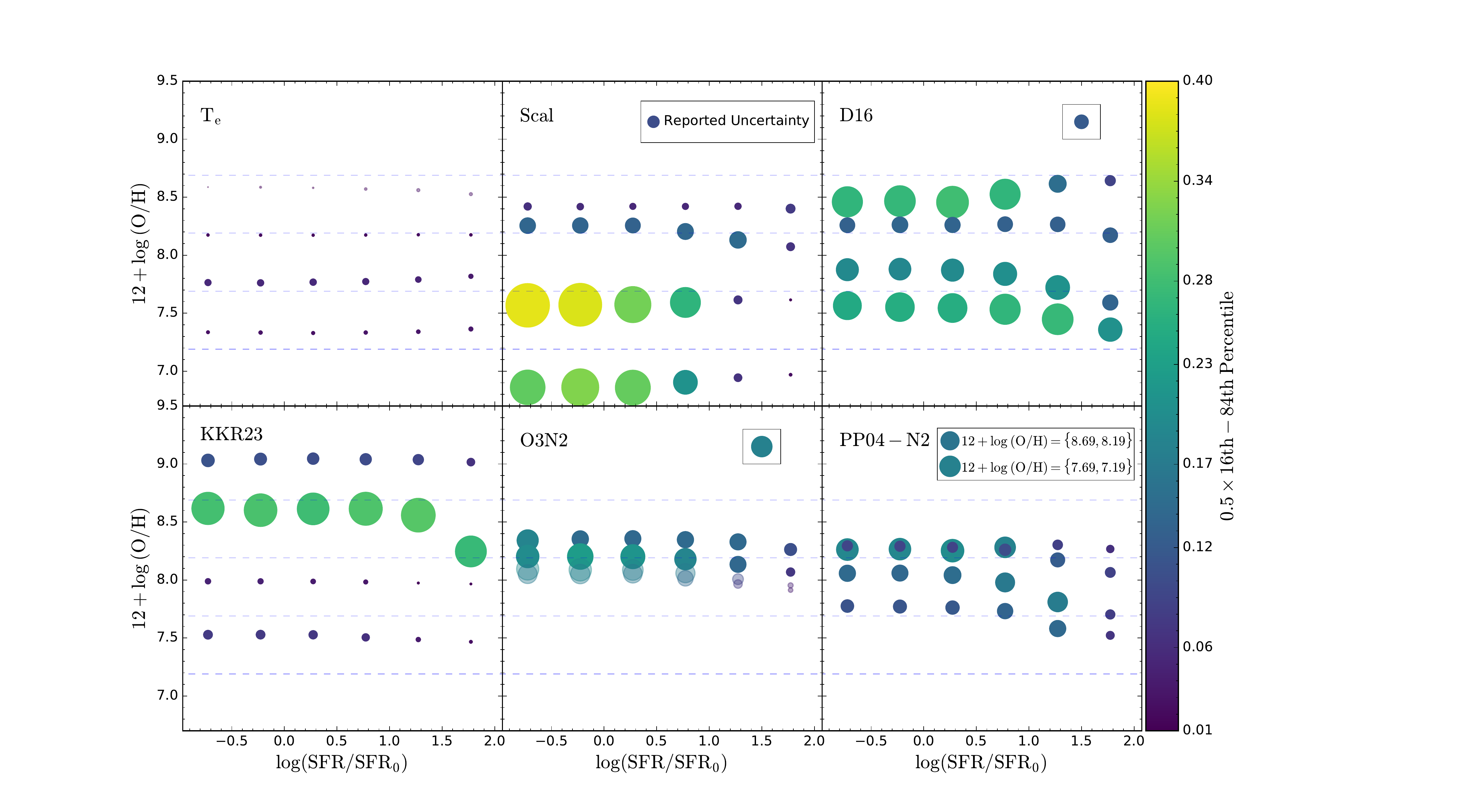}
     \caption{Same as \autoref{fig:variation_with_SFR_metallicity}, but using an observability limit of $10^{35}$ erg s$^{-1}$ in the H$\alpha$ line, rather than our fiducial $10^{36}$ erg s$^{-1}$.}
     \label{fig:variation_with_metallicity_SFR_cut_35}
 \end{figure*}

We show the results of this analysis in \autoref{fig:variation_with_metallicity_SFR_cut_35}. We can see from the figure that the trend of stochastic variation with metallicity and SFR remains largely the same as that shown in \autoref{fig:variation_with_SFR_metallicity}, but that the absolute spread in inferred metallicity increases for some diagnostics. In general, T$_{e}$, $\rm KKR_{23}$, and $\rm O3N2$ are unaffected by the observability limits. By contrast, $\rm D16$ and $\rm PP04-N2$ show a significant increase  ($> 0.1$ dex) in dispersion at solar metallicity, such that the dispersion at higher metallicity becomes comparable to that at low metallicities. For Scal, on the other hand, the already-high dispersion at lower metallicities and SFRs increases even more, reaching $0.4$ dex for the lowest SFR bin at $12+\log(\textrm{O/H}) = 7.89$. This is again much larger than the stated $\approx 0.1-0.15$ dex uncertainty of the diagnostic.

\section{Discussion}\label{sect:Disscusion}

So far we have explored how metallicities inferred from optical line diagnostics vary in response to changes in the degree of stochasticity ($\rm SFR/SFR_{0}$), the true metallicity of the underlying nebula ($Z_{\rm in}$), and the sensitivity of the observation. Depending on these parameters and on the choice of diagnostic, the stochasticity-induced spread can be as small as $\sim 0.01$~dex, or as large as $\sim 0.4$~dex. We now seek to understand the physical mechanisms that ultimately drive these fluctuations. We also try to separate out the two complementary effects that make up the stochasticity: inadequate sampling of the mass distribution and the age distribution of the stellar population. 

\subsection{Mechanisms for Stochastic Variation}

\begin{figure*}
    \centering
    \includegraphics[width = \textwidth]{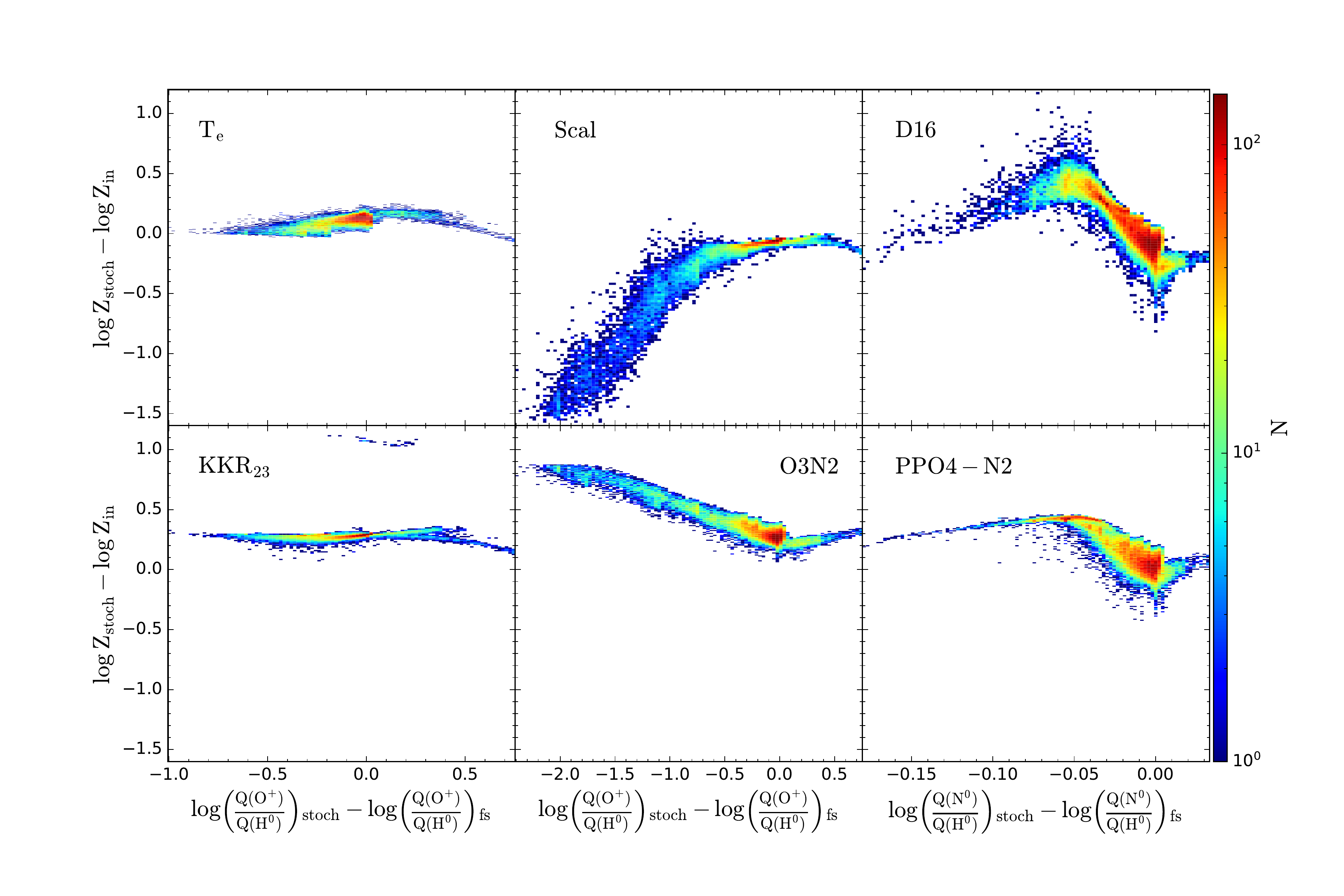}
    \caption{The joint distribution of metallicities and $Q\rm(X)$ values of stochastic runs ($\log Z_{\rm stoch}$) with true metallicity $\log Z/Z_\odot = -1.0$ and density in the range $n_{\rm II} = 10$--$100$~cm$^{-3}$. Each panel is for a different diagnostic, as indicated, and colours are proportional to the logarithm of the number of realisations in a particular bin in the $xy$-plane. The value on the vertical axis of each panel is the difference between the metallicity predicted by that diagnostic for a particular stochastic realisation, $Z_{\rm stoch}$, and the true metallicity. The horizontal axis is the ratio of luminosity in photons capable of ionising species X, $Q({\rm X})$, to luminosity of H$^0$-ionising photons, $Q({\rm H}^0)$, and we plot the value in a particular stochastic run relative to the value $[Q({\rm X})/Q({\rm H}^0)]_{\rm fs}$ for the case of a fully-sampled stellar population. The species X for which we plot is O$^+$ for the $T_e$, Scal, KKR$_{23}$ and O3N2 diagnostics, and N$^0$ for PP04-N2 and D16.}
    \label{fig:qx_with_metallicity}
\end{figure*}

The primary driver of stochastic variation in the inferred metallicity is variation in the ionising spectral shape, and in particular variations in the spectral shape that induce changes in the ionisation states of the metal atoms upon which metallicity diagnostics rely. To demonstrate this, we examine the quantity $Q({\rm X})$, as defined for $\rm X = H^{0}$ in \autoref{q_h}: the photon luminosity of the stellar population, counting only photons that are capable of ionising X to X$^{+}$. For the choice of species X, we consider the highest-energy ionisation transition relevant to the lines used by a given diagnostic -- thus for $T_e$, KKR$_{23}$, Scal, and O3N2 we compute $Q({\rm O}^+)$, since these diagnostics all use one or both components of the [O~\textsc{iii}] $\lambda4959,5007$ doublet, while for PP04-N2 and D16 we compute $Q({\rm N}^0)$, since these use [N~\textsc{ii}] $\lambda 6584$. We show the joint distribution of $Q({\rm X})$ and inferred metallicity from our stochastic runs, $\log Z_{\rm stoch}$, in \autoref{fig:qx_with_metallicity}. In this figure, we show one panel for each diagnostic, and in these panels, the vertical axis shows $\log Z_{\rm stoch} - \log Z_{\rm in}$ , the difference between the inferred and true metallicities, and the horizontal axis shows $\log [Q({\rm X})/Q({\rm H}^0)]_{\rm stoch} - \log [Q({\rm X})/Q({\rm H}^0)]_{\rm fs}$, where ``fs'' refers to the fully-sampled case; thus a value of 0 on the horizontal axis corresponds to a spectrum where the ratio of X-ionising photons to H$^0$ ionising photons is the same as in the fully-sampled case, and values larger (smaller) than zero indicate an excess (deficit) of such photons. Here, we show the subset of stochastic runs with ($\log Z_{\rm in}/Z_{\odot} = -1.0$) and with densities in the range $n_{\textsc{ii}} = 10$--$100$\,\si{cm^{-3}}. Note that this figure includes only runs where the emission would be observable, as described by our observability criteria in \autoref{subsect: observability_limits}.
 
We see that most of our runs are positioned close to zero in the horizontal direction, indicating, as expected, that the ratio $Q({\rm X})/Q({\rm H}^0)$ in our stochastic runs scatters about the value expected for the fully-sampled case. However, it is also clear that it is the runs that are farther away from the zero value on the x-axis that bring about the observed spread in inferred metallicity. That is, stochastic runs where the inferred metallicity is far from the correct value are preferentially those where the ionising spectral shape, as parameterised by $Q({\rm X})/Q({\rm H}^0)$, differs most strongly from that expected in the fully-sampled case. The case of O3N2 is particularly clear, and, thanks to its simplicity, provides a useful lens for understanding the origin of the effect. In this diagnostic, the inferred metallicity simply scales with the strength of the [O~\textsc{iii}] $\lambda 5007$ line as $\log({\rm O}/{\rm H}) \propto -0.32 \log([\mbox{O~\textsc{iii}}] \lambda 5007 / {\rm H}\beta)$.\footnote{This diagnostic also depends on the ratio $[{\rm N}~\textsc{ii}] \lambda 6584/{\rm H}\alpha$, but this varies by much less than the $[\mbox{O~\textsc{iii}}] \lambda 5007 / {\rm H}\beta$ ratio, so we can focus on the latter.} Thus, a 1~dex decrease in the luminosity of the [O~\textsc{iii}] line translates to an $\approx 1/3$~dex increase in inferred metallicity. Examining \autoref{fig:qx_with_metallicity}, we see that the error in the inferred metallicity as a function of $Q({\rm O}^+)/Q({\rm H}^0)$ shows precisely this behaviour: a 1~dex decrease in $Q({\rm O}^+)/Q({\rm H}^0)$ leads to an $\approx 1/3$~dex rise in $\log Z_{\rm stoch}$. The physical explanation is simply that the [O~\textsc{iii}] $\lambda 5007$ line comes from O$^{++}$ ions, which are produced by O$^+$-ionising photons; thus the ratio $[\mbox{O~\textsc{iii}}] \lambda 5007 / {\rm H}\beta$ is close to linearly proportional to the ratio $Q({\rm O}^+)/Q({\rm H}^0)$.

With this mechanism understood, we can now also understand the behaviour of other diagnostics. The $T_e$, and KKR$_{23}$ diagnostics are, for the case of input metallicity ($\log Z_{\rm in}/Z_{\odot} = -1.0$), much less vulnerable to stochasticity than O3N2 and Scal because they attempt to correct for the unknown ionisation state of oxygen in the nebula -- by directly measuring lines produced by both O$^+$ and O$^{++}$ ions (e.g both [O~\textsc{ii}] $\lambda 3727$ and [O~\textsc{ii}] $\lambda 5007$ ). Because they include an explicit correction for the ionisation state, these diagnostics are less easily fooled when the oxygen ionisation balance shifts unexpectedly in response to stochastic variations in the ionising spectrum. The $T_e$ and KKR$_{23}$ methods continue to work reasonably well even at metallicities close to the Solar metallicity. On the other hand, the Scal diagnostic attempts to correct for the unknown ionisation states of oxygen using lines produced by N$^+$ and S$^+$ as proxies for the hard-to-observe auroral lines of O$^+$. This works at solar metallicities but encounters problems at low metallicity because the N$^+$ and S$^+$ lines on which it relies cease to provide a good approximation to the unobserved O$^+$. Thus Scal is relatively immune to stochasticity at high metallicity, but not at low metallicity.

By contrast, O3N2, PP04-N2, and D16 do not include an implicit or explicit ionisation correction at all. For this reason, they can make substantial errors when the ionising spectrum contains a deficit or excess of O$^+$-ionising photons, unexpectedly shifting the ionisation balance in the nebula. This error is not strongly dependent on the underlying metallicity of the nebula.

\subsection{Mass Sampling vs Age Sampling}

\begin{figure*}
    \includegraphics[width=\textwidth]{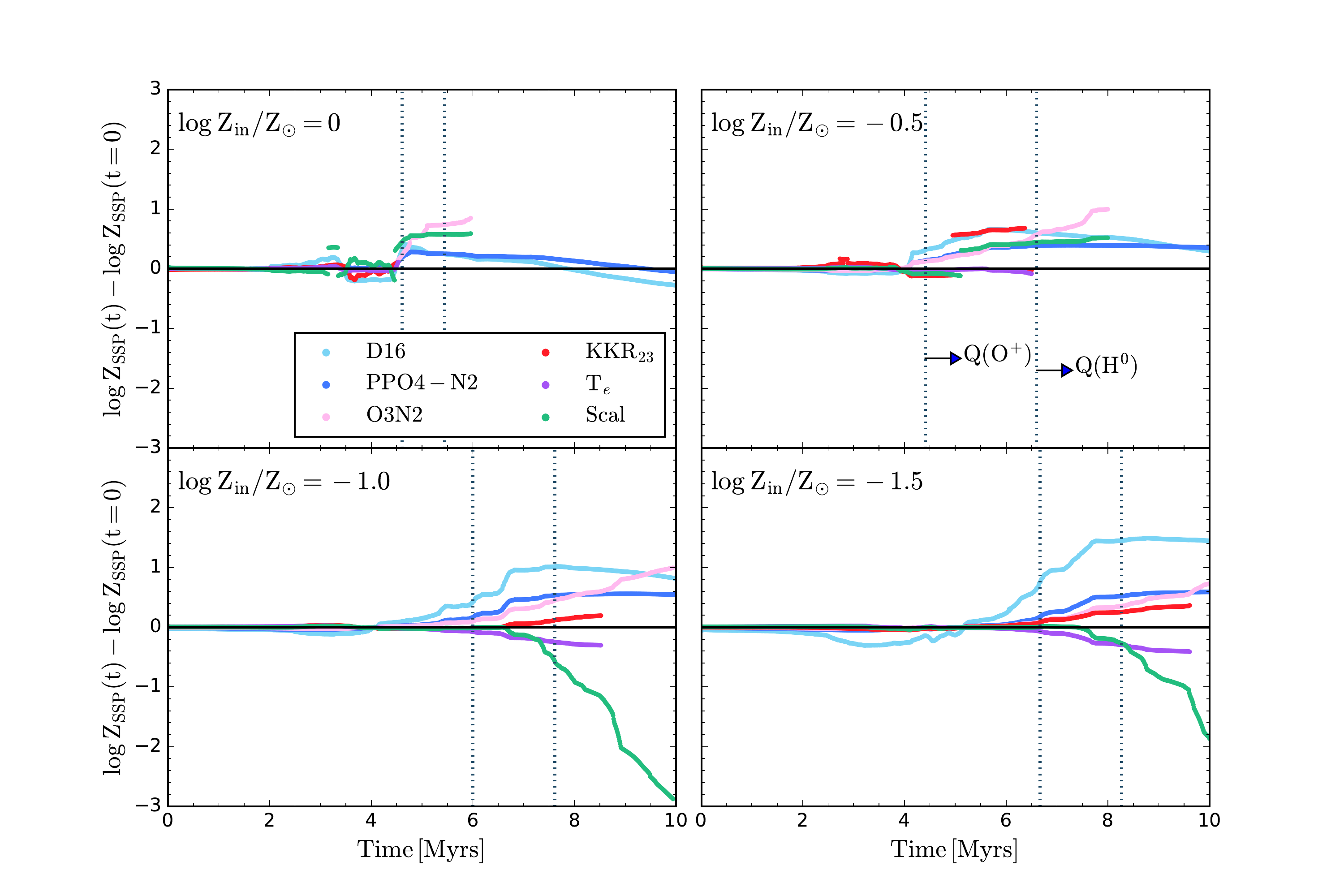}
    \caption{The evolution of the metallicity one would infer by applying a particular diagnostic to a fully-sampled SSP cluster of mass $10^{4}\rm \: M_{\odot}$; the quantity on the vertical axis is the difference between the metallicity $Z_{\rm SSP}(t)$ inferred for a SSP population of age $t$ and that inferred for a population of age zero. The first and the second vertical lines represent the time at which the  O$^+$- and H$^0$-ionising luminosities $\rm Q(O^{+})$ and $\rm Q(H^{0})$ fall to $10\%$ of their initial values, respectively.
    }
    \label{fig:delta_cluster}
\end{figure*}

\begin{figure}
    \centering
    \includegraphics[width = 0.47\textwidth]{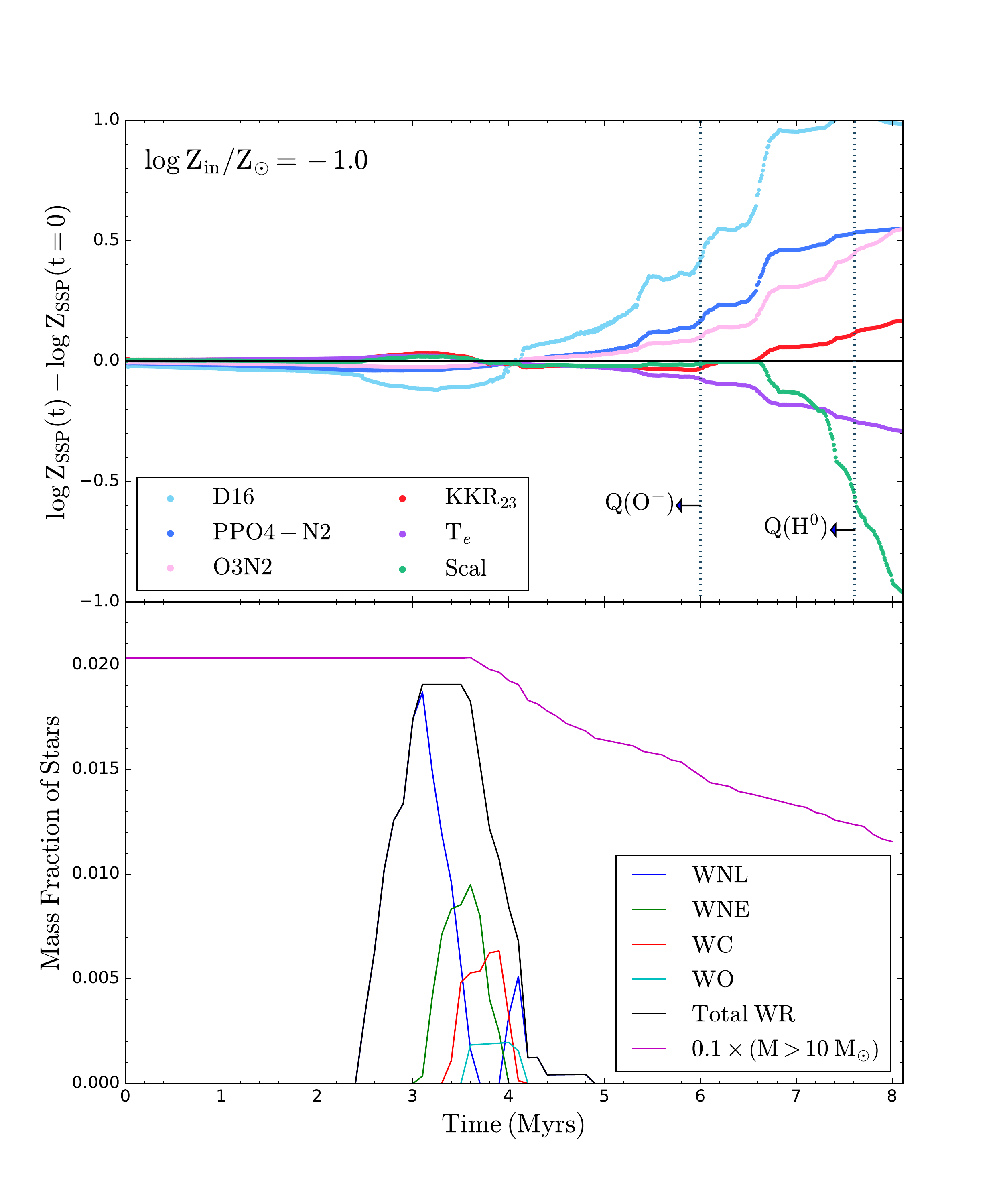}
    \caption{
    Top: inferred metallicity for a SSP as a function of age, using our six sample diagnostics, for a cluster of true metallicity $\log Z_{\rm in}/Z_{\odot} = -1.0$; this panel is identical to the lower left panel of \autoref{fig:delta_cluster}. Bottom: fraction of the stellar population by mass in various WR spectral classes; here WNL and WNE are late and early type WN stars, Total WR indicates the sum of all WR sub-types, and $0.1\times(M>10\,\mathrm{M}_\odot)$ indicates the total mass of all stars above $10$ M$_\odot$, reduced by a factor of 10 so that it is visible on the same scale.
    }
    \label{fig:massFraction_withTime}
\end{figure}

We now explore the relative importance of the two complementary factors that contribute to the spread in inferred metallicities: mass and age sampling. For this purpose, we compare the spread in inferred metallicites for a stellar population that is fully sampled in mass but not in age (the SSP case) with the fully stochastic runs that are not fully-sampled in either of the two. In order to explore how imperfect sampling in age affects metallicity diagnostics, we plot the metallicity derived by applying each diagnostic to an SSP run as a function of age in \autoref{fig:delta_cluster}. In the figure, each panel shows SSP runs at true metallicity values $\log Z_{\rm in}/Z_{\odot} = -1.5, -1.0, -0.5, 0$, as indicated in the legend. The horizontal axis shows the age and the vertical axis shows $\log Z_{\rm SSP} (t)  - \log Z_{\rm SSP} (t = 0)$, i.e, the difference between the metallicity inferred from the population at age $t$ and one of age 0. Lines of a given diagnostic end at an age when it is no longer possible to measure the metallicity for a SSP of total mass $10^{4}$~$M_{\odot}$ using that diagnostic, based on our adopted observability limits\footnote{For the SSP, we use only the relative line cut-offs because the absolute line luminosities can be scaled by increasing the cluster mass.} (see \autoref{subsect: observability_limits}). The dotted vertical lines show the ages at which the $Q\rm (H^{0})$ and $Q\rm (O^{+})$ fall to 10$\%$  of their initial values ($t = 0$).

In \autoref{fig:delta_cluster}, we can immediately see that the inferred metallicities for all diagnostics stay nearly constant at ages $\lesssim 3$--$5$~Myr, depending upon the metallicity, before drifting away form their initial values at older ages. The initial drift away from zero and the subsequent rise in the uncertainty at later times can be associated with the Wolf-Rayet and massive stars. We can see this explicitly in \autoref{fig:massFraction_withTime}, where we plot the fraction of the population (by mass) in WR stellar classes as a function of SSP age for example case $\log Z_{\rm in}/Z_{\odot} = -1.0$. We construct this plot using the WR assignments provided in the MIST evolutionary tracks that we use throughout this work \citep{mist_2016}. As is clear from the plot, the onset of the small fluctuations in $\log Z_{\rm SSP}(t)$ coincides with the initial appearance of WR stars. However, the rapid growth in fluctuations coincides with the disappearance of the Wolf-Rayets and massive stars from the population. It is straightforward to understand why: the first appearance of Wolf-Rayet stars significantly alters the ionising spectral shape by increasing the hard ionising photon flux. A few Myr later these stars begin to die off, and there is a rapid drop in $Q({\rm O}^+)$. The resulting fluctuations in inferred metallicity are in opposite directions, depending on whether a particular diagnostic increases or decreases as the nebular ionisation state changes. Clearly the appearance of WR stars, more importantly, their disappearance as a function of age is capable of inducing significant shifts in the metallicity derived by at least some diagnostics, even for stellar populations that fully sample the IMF. 

In order to quantify the importance of partial IMF sampling versus age sampling, we next investigate how much of the stochastic variations result from inadequate age sampling alone. For this purpose, we generate a stochastic-in-time library of models by randomly selecting SSP models at different ages, from age 0 to the age at which $Q({\rm H}^0)$ falls to 10\% of its initial value (denoted by the second vertical line in \autoref{fig:delta_cluster}), which we take to be the lifetime of the ionising stars. This library represents what we would observe if all stars were formed in SSP clusters, and each spaxel in an IFU observation contained exactly one such cluster, but each of these clusters perfectly sampled the IMF. For each of these libraries, we have a distribution of inferred metallicities for each diagnostic (as usual, discarding cases where the lines for a particular diagnostic would be too faint to see), and we can compute the 16th to 84th percentile range of the resulting inferred metallicities, which we denote $\Delta Z_{\rm SSP}$. We can then compare this range to the 16th - 84th percentile range produced by our fully-stochastic runs (or some subset thereof), $\Delta Z_{\rm stoch}$, which are stochastic in both age and stellar mass.

\begin{figure*}
    \centering
    \includegraphics[width = \textwidth]{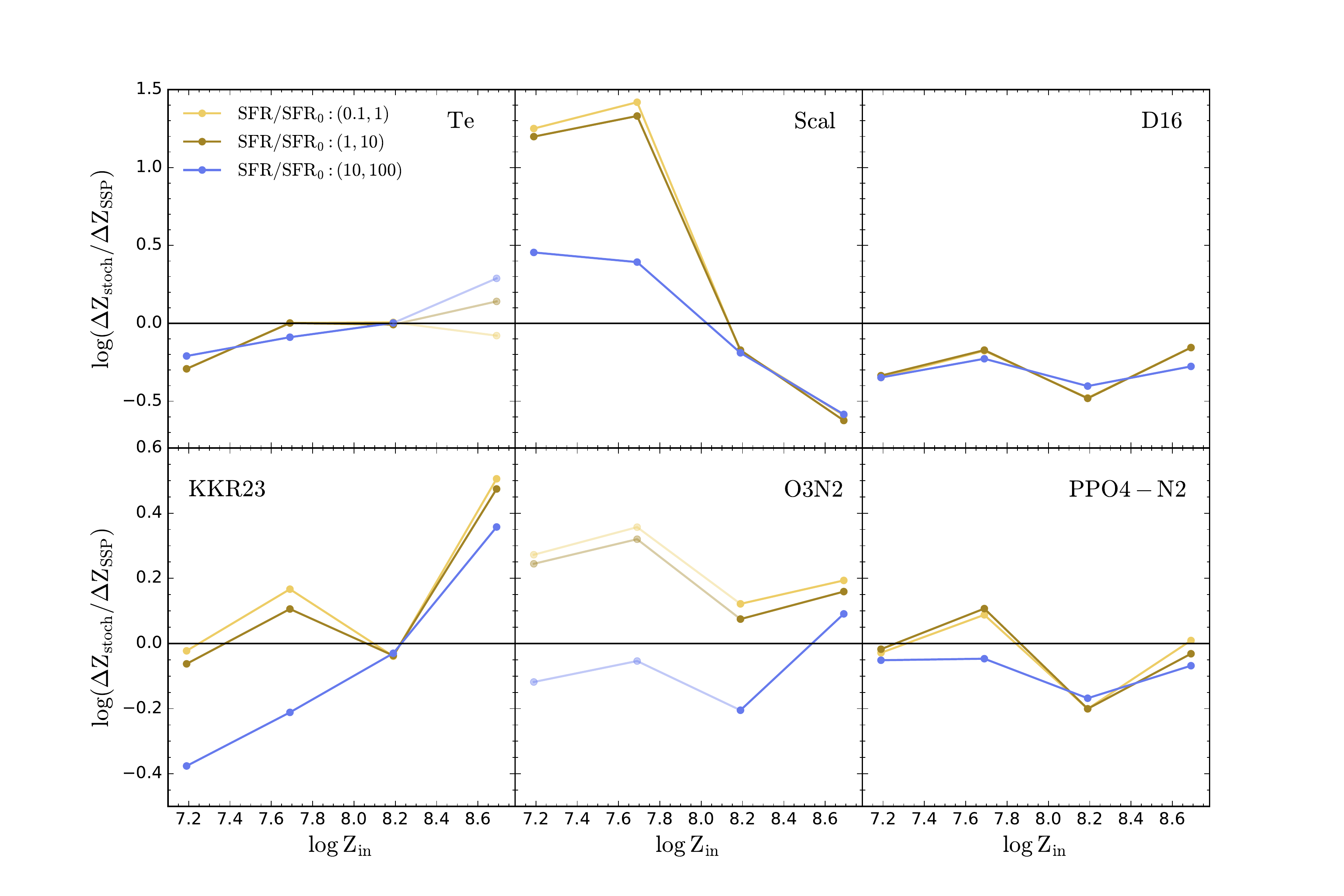}
    \caption{Each panel, corresponding to the diagnostic indicated, shows the ratio of the 16th to 84th percentile range for metallicities inferred from fully stochastic runs, $\Delta Z_{\rm stoch}$, divided by the corresponding range in metallicities inferred for runs where the stellar population consists of a SSP cluster that is fully sampled in mass, $\Delta Z_{\rm SSP}$. We plot this quantity as a function of true metallicity $Z_{\rm in}$. Colours indicate stochastic runs in different bins of $\rm SFR/SFR_{0}$ from $0.1$--$1$, $1$--$10$, and $10$--$100$, as indicated in the legend. The faded points represent the regions where the diagnostic is not reliable - either due to faint lines or because of it was not calibrated in that metallicity range.
    }
    \label{fig:monoAge_percentiles}
\end{figure*}

We compare the uncertainties observed in the SSP and fully stochastic runs in \autoref{fig:monoAge_percentiles}. In the figure each panel represents a metallicity diagnostic. On the vertical axis we plot $\Delta Z_{\rm stoch}/\Delta Z_{\rm SSP}$, while on the horizontal axis we plot the true metallicity ($Z_{\rm in}$) of our models. Note that the stochastic runs for which we compute $\Delta Z_{\rm stoch}$ are those with densities in the range $n_{\rm II} = 10$--$100$ \si{cm^{-3}} and that we have divided the stochastic runs into three bins of ${\rm SFR}/{\rm SFR}_0$, running from $0.1$--$1$, $1$--$10$, and $10$--$100$. 

The first thing that we notice is that the blue lines, indicating fully stochastic runs with ${\rm SFR}/{\rm SFR}_0 = 10$--$100$, generally show a spread $\Delta Z_{\rm stoch}$ that is quite close to $\Delta Z_{\rm SSP}$, i.e., $\Delta Z_{\rm stoch} \approx \Delta Z_{\rm SSP}$ to within $\approx 50\%$ for most diagnostics at most metallicities (one exception being KKR$_{23}$). This means that the stochasticity present in our high-SFR runs is mostly a result of imperfect time sampling. A simple physical interpretation is that, at a relatively high SFR, we have enough stars to sample the range of possible stellar masses reasonably well. However, because stars are clustered in time (present in SSP clusters), not all stages of stellar evolution are present in any given patch -- indeed, this is consistent with our earlier estimate that, even at $\mbox{SFR}/\mbox{SFR}_0 \approx 30$, we expect a given galaxy patch to contain only one cluster with age $<4$~Myr that is massive enough to fully sample the IMF. Consequently, stochasticity in time dominates in this range of star formation rate.

Now, looking at different diagnostics separately, we see for the diagnostics that do not use the oxygen lines at all -- D16 and PP04-N2 -- the uncertainties in stochastic runs and SSP runs are comparable. Thus, in this case, inadequate time-sampling is the major contributor to the uncertainty. In contrast, for the diagnostics that use the $\left [\rm O \textsc{ii} \right ]$ and/or $\left [\rm O \textsc{iii} \right ]$ lines directly -- $T_e$, Scal, KKR$_{23}$ and O3N2 -- we see that mass-sampling plays a major role at lower SFRs, because $\Delta Z_{\rm stoch}$ is generally larger than $\Delta Z_{\rm SSP}$ for ${\rm SFR}/{\rm SFR}_0 < 10$. We tentatively interpret this as indicating the importance of WR stars, which are the only stars capable of producing appreciable numbers of photons capable of ionising O$^+$ to O$^{++}$. At any given age, only stars born within a quite narrow range of mass will be in their WR phase. Consequently, the stochasticity in mass as well as in age is significant for diagnostics that rely on the presence of O$^+$-ionising photons.

\section{Summary}
\label{sect:conclusion}

In this paper we quantify how the inferred metallicities of $\rm H~\textsc{ii}$ regions are affected by stochastic variations in the stellar populations that drive them. More precisely, we explore to what extent metallicities inferred using a range of diagnostics fluctuate in response to stochastic fluctuations in the shape of the ionising continuum produced when the stars driving the H~\textsc{ii} region are not fully-sampled in mass (IMF) and in stellar age. Consideration of stochastic effects is important for the interpretation of modern IFU observations, because these now routinely sample $\sim 100$~pc-sized or even smaller patches in nearby galaxies. In such a small region, there are unlikely to be enough massive stars to represent a fully-sampled stellar population, but the impact of this imperfect sampling on the metallicities that we derive has not previously been subject to extensive systematic exploration.

We use the \textsc{slug} code to generate stochastic realisations of stellar populations in patches of galaxies. We then use the photoionisation code \textsc{cloudy} to calculate the line emission from nebulae driven by these stochastic stellar populations, and feed the resulting line emission into six of the most commonly used metallicity diagnostics -- $T_e$, KKR$_{23}$, Scal, O3N2, PP04-N2 and D16 -- to derive inferred metallicities. We repeat this process for $\sim 1$~million stochastic realisations where we vary the star formation rate, metallicity, and ionised gas density of our $\rm H~\textsc{ii}$ regions.

We find that the primary channel through which stochasticity affects metallicity inference is via the variations it causes in the highest energy parts of the ionising spectrum, which are primarily produced by Wolf-Rayets stars, a phase of stellar evolution that is both short in duration and narrow in mass range, and thus is subject to considerable stochastic fluctuation. We have shown this results in a deficit in the high-energy part of the ionising spectrum that directly drives a spread in metallicities that we infer by applying line diagnostics to nebular emission. We find that the size of this spread depends on the choice of diagnostic, the true metallicity (which affects both the stellar spectra and the ionisation and temperature balance of the nebula), the star formation rate (which dictates how many massive stars are present, and thus the size of fluctuations in the spectral shape), and the survey sensitivity limit (which determines at what point low-mass, poorly-sampled H~\textsc{ii} regions become so dim that we cease to observe them at all). In the most stochastic cases we consider, $\sim 100$~pc-sized galaxy patches with star formation rates from 10 to 100\% of the value found in the Solar neighbourhood, we observe spreads in the inferred metallicity from some of the diagnostics that can be as high as $\sim 0.3$~dex for sensivity limits comparable to those of the deepest existing IFU surveys, rising to $\sim 0.4$ dex for hypothetical future surveys going a factor of 10 deeper. We see that at this point, the uncertainties due to the stochastic fluctuations, at times,  dominate the reported uncertainties. The spread drops to $\sim 0.1$~dex as we increase the star formation rate up to $100$ times that in the Solar neighbourhood and sample the age and mass distributions more adequately.

However, we also find that some diagnostics correct for stochastic effects explicitly and are thus insensitive to stochasticity, regardless of its size. The key distinction between diagnostics that are stochasticity-resistant and those that are not is the extent to which the diagnostic samples more than one ionisation state of the ions being used. Diagnostics such as the electron temperature method ($T_e$) or KKR$_{23}$, which make use of lines from both O$^+$ and O$^{++}$, for example, show stochastic variations in the derived metallicity $\lesssim 0.2$~dex, regardless of the star formation rate of the regions being sampled. By contrast, diagnostics such as O3N2 that rely on a single ionisation state of O have stochastic variations that depend considerably on the star formation rate.

We further show that the primary driver of variations at the highest star formation rates is stochastic sampling in age rather than in mass -- that is, even though a star formation rate $\sim 30$ times that found in the Solar neighbourhood is enough to guarantee a wide sampling of stellar masses, stochastic variations of $\sim 0.1-0.2$~dex in inferred metallicities still persist, because such a region is likely to contain only a single star cluster massive enough to fully sample the IMF, and thus will not contain all possible stellar ages. 

Our findings present a challenge for future surveys. The best protection against stochasticity is to use diagnostics that sample multiple ionisation states. Of the diagnostics we explore, $T_{e}$ performs best, with KKR$_{23}$ coming close. However, an important caveat to note is that the auroral lines used by these diagnostics are quite faint, and thus are hard to observe in regions where the star formation rate is low enough for stochasticity to be a concern in the first place. Scal, which avoids the faintest auroral lines and uses brighter proxies to estimate the ionisation state, shows strong resistance to stochastic variation at high metallicity, and thus is a good choice for Solar or near-Solar targets, but its S$_2$-based estimate of the ionisation state becomes subject to very large stochastic variations at metallicities $\lesssim 10\%$ of Solar. The strong-line diagnostics O3N2, PP04-N2, and D16 use lines that are likely to be detectable even in regions of low star formation rate, but are subject to quite large stochastic uncertainties, and thus the choice of relying on these methods versus more stochasticity-resistant ones that require weaker lines creates a trade-off between the number of regions that one can observe and the robustness of the derived metallicities. With the advent of IFUs and the high-resolution mapping of nearby galaxies that they enable, however, we can no longer ignore the trade-offs imposed by stochasticity, and we must take into account stochastic uncertainties when we interpret observations of metallicity trends. This study can be used as a reference for both of these purposes.

\section*{Data availability}

The outputs of the \textsc{slug} and \textsc{cloudy} simulations upon which this paper is based are available for download from \url{https://cloudstor.aarnet.edu.au/plus/s/x2yjsHdKkRjzgWD}. The data can be read and processed using the \textsc{slugpy} library distributed as part of the \textsc{slug} package, available from \url{https://bitbucket.org/krumholz/slug2/src/master/}.

\section*{Acknowledgements}

MRK acknowledges support from an Australian Research Council (ARC) Future Fellowship (award FT180100375). CF acknowledges support from the ARC Discovery Projects scheme (grant DP170100603) and the Future Fellowship scheme (grant FT180100495). This research was undertaken with the assistance of resources and services from the National Computational Infrastructure (projects jh2 and ek9), which is supported by the Australian Government.




\bibliographystyle{mnras}
\bibliography{paper} 




\appendix

\section{Effects of gas density}\label{appendix}

 \begin{figure*}
    \hfill
     \centering
     \includegraphics[width =\textwidth]{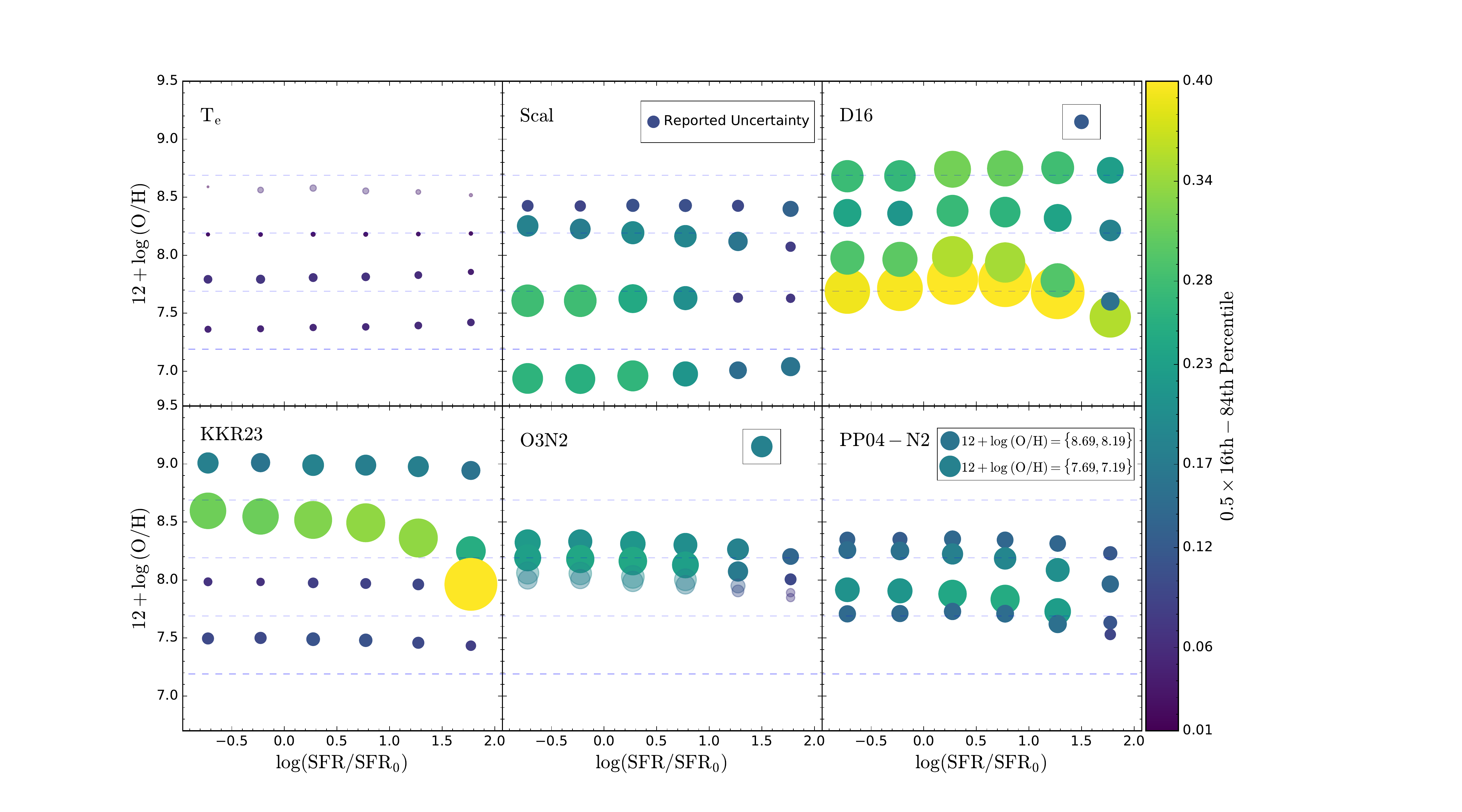}
     \caption{ 
     The same as
     \autoref{fig:variation_with_SFR_metallicity}, but for a range of H~\textsc{ii} region densities $n_{\textsc{ii}} = 10 - 10^4 $~\si{cm^{-3}}, uniformly sampled in $\log n_{\textsc{ii}}$.}
     \label{fig:variation_with_zin_allHden}
 \end{figure*}

Here, we explore the dependence of inferred metallicities on ionised gas number density. To put this discussion in context, we note that most observed $\rm H~\textsc{ii}$ regions have densities in the range $\sim 10$--$100$~\si{cm^{-3}} \citep{tremblin_2014}. Thus, we would typically expect an IFU survey of a galactic disc to contain only a handful of spaxels for which $n_{\textsc{ii}}\geq 1000$~\si{cm^{-3}}. To examine how stochasticity is likely to affect these spaxels, we repeat the analysis presented in \autoref{subsect: Star Formation Rate} for the full density range ($10$--$10^4$~\si{cm^{-3}}) present in our simulation library. This almost certainly exaggerates the effects of density, since our simulation library is uniformly-sampled in $\log \: n_{\textsc{ii}}$, making high-density regions far more common than is likely to be the case in reality.

We show the results in \autoref{fig:variation_with_zin_allHden}; other than the density range used to generate it, this figure is identical to \autoref{fig:variation_with_SFR_metallicity}. Comparing the two figures, we immediately notice an increase in the uncertainty range of the $\rm D16$ diagnostic across the whole metallicity range. Scal is the next-most affected, though by much less than D16, and all the other diagnostics show negligible changes ($\lesssim 0.1$~dex) when compared to our analysis of the models with a narrower density range. Thus, we see that even when we include the  \hii~regions with ionised gas densities far from the commonly observed values, we do not see major changes in our results and conclusions. The reason the D16 diagnostic is affected so strongly is that [S~\textsc{ii}] $\lambda6717,31$ doublet upon which it relies has a relatively low critical density, and thus can be collisionally-quenched in the higher-density H~\textsc{ii} regions we are now sampling. Scal, which also uses this doublet, is affected as well, but less severely.


\bsp	
\label{lastpage}
\end{document}